\documentclass[fleqn,usenatbib]{mnras}

\usepackage[T1]{fontenc}
\usepackage{ae,aecompl}

\DeclareRobustCommand{\VAN}[3]{#2}
\let\VANthebibliography\thebibliography
\def\thebibliography{\DeclareRobustCommand{\VAN}[3]{##3}\VANthebibliography}

\usepackage{graphicx}	
\usepackage{amsmath}	
\usepackage{amssymb}	

\title[MZR of ETGs from VANDELS]{The star formation history and the nature of the
mass-metallicity relation of passive galaxies at 1.0$<$$z$$<$1.4 from VANDELS.
}
\author[P. Saracco et al.]{P. Saracco,$^{1}$\thanks{E-mail: paolo.saracco@inaf.it},
F. La Barbera$^{2}$, R. De Propris,$^{3,1,10}$, D. Bevacqua$^{4,1}$, D. Marchesini$^5$, 
\newauthor G. De Lucia$^6$, F. Fontanot$^6$, M. Hirschmann$^{7,6}$, M. Nonino$^6$, A. Pasquali$^8$, \newauthor C. Spiniello$^{9,2}$, C. Tortora$^2$ 
\\
$^{1}$INAF - Osservatorio Astronomico di Brera, via Brera 28, 20121 Milano, Italy\\
$^{2}$INAF - Osservatorio Astronomico di Capodimonte, sal. Moiariello 16, 80131 Napoli, Italy\\
$^{3}$FINCA, University of Turku, Vesilinnantie 5, 20014 Turku, Finland\\
$^{4}$Universita\'\ degli studi dell'Insubria, via Valleggio 11, 22100 Como, Italy\\
$^{5}$Tufts University, Physics and Astronomy Department, 574 Boston Ave, Medford, 02155 MA, USA\\
$^{6}$INAF - Osservatorio Astronomico di Trieste, via G.B. Tiepolo 11, 34143 Trieste, Italy\\
$^{7,6}$ Institute of Physics, Laboratory for Galaxy Evolution, Ecole Polytechnique Federale de Lausanne, Observatoire de Sauverny, \\ Chemin Pegasi 51, 1290 Versoix, Switzerland \\
$^{8}$Astronomisches Rechen-Institut, Zentrum f\"ur Astronomie der Universit\"at Heidelberg,
M\"onchhofstrasse 12 - 14, 69120 Heidelberg, Germany \\
$^{9}$Sub-Dep. of Astrophysics, Dep. of Physics, University of Oxford, Denys Wilkinson Building, Keble Road, Oxford OX1 3RH, UK\\
$^{10}$Department of Physics and Astronomy, Botswana International University of Science and Technology, Private Bag 16, Palapye, Botswana }

\date{Accepted 2023 January 18. Received 2023 January 11; in original form 2022 November 04}

\pubyear{2023}

\begin{document}
\label{firstpage}
\pagerange{\pageref{firstpage}--\pageref{lastpage}}
\maketitle

\begin{abstract}
We derived stellar ages and metallicities [Z/H] for $\sim$70 
passive early type galaxies (ETGs) selected from VANDELS  survey over the
redshift range 1.0$<$$z$$<$1.4 and stellar mass range 10$<$log(M$_*$/M$_\odot$)$<$11.6.
We find significant systematics in their estimates depending on models
and wavelength ranges considered.
Using the full-spectrum fitting technique, we find that both [Z/H] and age increase with 
mass as for local ETGs.
Age and metallicity sensitive spectral indices independently confirm these trends. 
{According to EMILES models,} for 67 per cent of the galaxies we find [Z/H]$>$0.0, a percentage which rises to $\sim$90
per cent for log(M$_*$/M$_\odot$)$>$11 where the mean metallicity is [Z/H]=0.17$\pm$0.1.
A comparison with homogeneous measurements at similar and lower redshift 
does not show any metallicity evolution over the redshift range 0.0$<$$z$$<$1.4.
The derived star formation (SF) histories show that the stellar mass fraction
formed at early epoch increases with the mass of the galaxy.
Galaxies with log(M$_*$/M$_\odot$)$>$11.0 host stellar populations with [Z/H]$>$0.05,
formed over short timescales ($\Delta{t50}$$<$1 Gyr) at early epochs (t$_{form}$$<$2 Gyr),
implying high star formation rates (SFR$>$100 M$_\odot$/yr) in high mass density 
regions (log($\Sigma_{1kpc}$)$>$10 M$_\odot$/kpc$^2$).
This sharp picture tends to blur at lower masses: log(M$_*$/M$_\odot$)$\sim$10.6 
 galaxies can host either old stars with [Z/H]$<$0.0 or younger stars with [Z/H]$>$0.0, depending 
 on the duration ($\Delta{t50}$) of the SF.
The relations between galaxy mass, age and metallicities are therefore
 largely set up {\it ab initio} as part of the galaxy formation process.
{Mass, SFR and SF time-scale all contribute to shape up the stellar mass-metallicity
relation with the mass that modulates metals retention.}
\end{abstract}

\begin{keywords}
galaxies: evolution; galaxies: elliptical and lenticular, cD;
           galaxies: formation; galaxies: high redshift
\end{keywords}



\section{Introduction}
Early-type galaxies (hereafter ETGs) are known to obey a series of `scaling relations' with  small intrinsic scatter, 
most of them being a projection of a thin distribution in the multi
dimensional fundamental plane \citep[FP, e.g.,][]{djorgovski87, dressler87,jorgensen96}. 
In particular,
the colour-magnitude (CM) relation \citep{Sandage1977,Bower1992} is observed to
persist, with largely unchanged slope and scatter, except for the effects of stellar aging, even at $z=2$ \citep{Blakeslee2003,Mei2006a,Mei2006b,Mei2012,newman14,Lemaux2019,Willis2020}. This is conventionally interpreted as a relation between mass and metallicity, with the intrinsic scatter being provided by a spread in age \citep{Kodama1997,gallazzi06}. 
Spectroscopy confirms that the colour-magnitude relation 
is driven by a mass-metallicity trend out to $z=1.2$ at least
\cite[e.g.,][]{jorgensen17,Lemaux2019,saracco19}. 
However, interpretation of galaxy colours and their evolution in terms of age and metallicity is hampered by the well-known degeneracies between these two quantities \citep[e.g.][]{worthey1994}.

The simplest interpretation is one where the CM relation is established by 
an intense burst of star formation (SF) at high redshift (e.g, \citealt{Pipino2004,dekel09}) 
where most of the stellar mass is formed {\it in situ} and the depth of the potential well determines the epoch at which stellar winds and supernova explosions manage to eject the remaining gas from the forming proto-galaxy. 
{ However, it is not necessary to form stars {\it in situ} to have older stars in more massive galaxies.
In fact, older stars can also be found in more massive galaxies when later assembled through
merging
\citep[e.g.,][]{delucia04}.
Moreover, integrated star formation histories (SFHs) for stars belonging to massive galaxies can also
produce the correct scaling relation in models where mergers are significant
in the stellar mass assembly \citep[see, e.g., Fig.7 in][]{fontanot17}.} 

The age- and stellar metallicity-mass relations (MZR) are well studied in the local Universe
\citep[e.g.,][]{thomas05,gallazzi05,thomas10,choi14,mcdermid15}.
These studies established that most massive ETGs host, on average, 
older and more metal-rich stellar populations than lower mass galaxies, 
a result which is very counter-intuitive, even in the hypothesis of mergers, 
as time is needed to produce metals.

The metal content of ETGs and its evolution across time provides information 
about their past SF activity, their quenching phase and their evolution.
The total metallicity is the result of the duration of the SF and the gas exchange with
the inter/circumgalactic medium.
{ For instance, we expect that metallicities will be low and show no evolution with 
redshift if gas is quickly removed by an outflow (e.g., AGN-driven).
Indeed, if star formation is interrupted at early times due to the sudden removal of the gas
instead of being smoothly quenched at later times, the total metallicity will be lower 
\citep[e.g.,][]{delucia17,okamoto17,trussler20}. 
On the contrary, if the external gas supply (inflow) is stopped
or suppressed  by feedback mechanisms (e.g., AGN or SF feedback), or gas infall is 
negligible with 
respect to the rate at which gas is converted into stars (SFR),
the system resembles a closed box and metallicity increases rapidly to its maximum value 
\citep[e.g.][]{vazdekis97, peng15}.}

{ Since ETGs are seen to be quiescent,  it is expected that after the formation of the 
bulk of the galaxy stellar mass, their stellar population properties do not change 
significantly because of $in-situ$ SF.}
 If one then observes evolution in the metallicity of passive ETGs with redshift, 
 it must be due to progenitor bias \citep{vandokkum08,carollo13} or mergers adding 
 stellar populations to the mixture.
Minor mergers are expected to lower the metallicity over time (if the mass accreted is significant) since low-mass galaxies have lower metallicities; 
major mergers to leave the metallicity unchanged, since equal mass galaxies are
expected to have similar metallicity.
However, in the simulations, mergers (e.g., see \citealt{Mo:book}) require some fine tuning to match the scatter observed
in the scaling relations \citep[e.g.,][]{nipoti09,skelton12}. 
In general, if galaxies undergo numerous mergers, the scatter in any pre-existing 
relation between mass and metallicity 
is likely to increase, if it is even able to survive, as most mergers will take place between random objects.
In case of progenitor bias, the evolution (if any) depends on the formation and 
quenching mechanisms mentioned above.  

A better understanding of the star formation history (SFH) and
the mass assembly history of ETGs can be achieved by studying the stellar population
properties (age and metallicity) and their relationship with mass and other 
properties of galaxies at increasingly higher redshift.
\cite{gallazzi14} find relationships of increasing age and metallicity with the galaxy
mass at $z$$\sim$0.7 similar to those for ETGs in the local Universe,
consistently with the results of \cite{choi14}.
{ Similar non evolving trends between stellar metallicity and mass up to $z$$\sim$1.0
are found by \cite{ferreras09}.}
On the other hand, \cite{beverage21}, using LEGA-C data \citep{vanderwel16,vanderwel21},
find that galaxies at $z$$\sim$0.7 have metallicities 0.2dex lower than 
their local counterparts,
{ and that older galaxies have lower metallicities than younger ones}.

Stellar metallicity measurements at $z$$\ge$1 have been carried out 
for few massive or stacked galaxies \citep[e.g.][]{lonoce14,onodera15,saracco19,kriek19}
and a handful of galaxies at even higher redshifts ($z$$\sim$2.1, \citealt{kriek16};
$z$$\sim$3.35, \citealt{saracco20apj}).
Recently, \cite{carnall19,carnall22} used VANDELS data \citep{mclure18, pentericci18} to study
the metallicity of passive galaxies at $z$$\sim1.2$ and its evolution.
We will discuss and compare their results in \S\S 4 and 8.

Here we present the study of a sample of field early-type and passive galaxies at 
1.0$<$$z$$<$1.4 selected from the VANDELS survey data. 
We describe the dataset in the next section, where we also give details on our selection procedures and adopted stellar population models. 
In \S 3 we describe the method used to derive the stellar age and metallicity 
of galaxies, and we study the dependence on models and fitting assumptions.
In \S 4 we derive the stellar mass-metallicity relation (MZR) at $z$$\sim$1.2, study its evolution down to $z$$\sim$0 and we compare our results with the literature.
In \S 5 we derive the SFH of VANDELS galaxies.
In \S 6 we study the relationships between SFH and metallicity.
Section 7 summarizes the results which are then discussed in \S 8  where we present
also our conclusions.
In the Appendix \ref{sec:indices}, we describe the procedure used to measure absorption line spectral 
indices in the rest-frame wavelength range [2600-4350] \AA, and report them for the 
whole sample. 
Moreover, we use indices to test the results in a nearly model-independent way and mid-UV indices 
to constrain chemical abundances.

Throughout this paper we use a cosmology with
$H_0=70$ km s$^{-1}$ Mpc$^{-1}$, $\Omega_m=0.3$, and $\Omega_\Lambda=0.7$
and assume a \cite{chabrier03} initial stellar mass function (IMF).
Magnitudes are in the AB system, unless otherwise specified.

\section{Data and models}
\subsection{VANDELS observations and data}
A complete and detailed description of the VANDELS survey, spectroscopic observations, data
reduction and quality of the data can be found in \cite{mclure18,pentericci18} and \cite{garilli21}.
{  Fully reduced spectra have been made publicly available by the VANDELS team 
(see Sec. Data availability).
Here, we provide only a brief summary of the relevant points.

VANDELS is a VIMOS/VLT deep public spectroscopic survey in the wavelength range [4800--9800] \AA, of galaxies selected in the Hubble Space Telescope CANDELS and UDS fields.
The survey aims to detect star forming galaxies at high redshift and passive galaxies at 
$1.0 < z < 2.5$ with spectra of sufficient quality and resolution (R$\sim$600, 
FWHM$\sim$15 \AA\ at 9000\AA, 1 arcsec of slit width) to determine not only the redshift 
but also some stellar population parameters.

Spectra have an average dispersion of 2.5 \AA/pix.
The pixel scale of the images is 0.205''/pix.
The seeing, as measured on the science images, was below 1 arcsec 
in $\sim$90\% of the observations with a median value of $\sim$0.7''
\citep{garilli21},
corresponding to a spatial scale of $\sim$6 kpc at $z$$\sim$1.2.
Therefore, any possible radial variation of stellar population properties 
on angular scale lower than 0.7-1 arcsec, i.e. lower than ~6-8 kpc at z=1.2,
cannot be seen. 
It is worth noting that this angular scale is larger than the angular 
diameter (2R$_e$) of 95\% of our selected passive galaxies (see below) whose 
median value is 2R$_e$$\sim$3.5 kpc.

1D spectra were extracted applying the Horne optimal
extraction algorithm \citep{horne86} which delivers the maximum possible 
signal-to-noise ratio for each spectrum. 
This implies that spectra have been extracted within a variable aperture.
However, being the observations seeing limited, for the reasons discussed above,
the optimal extraction algorithm is not expected to introduce 
any systematic effect in the estimate of stellar population properties.
}

\subsection{Sample selection}
\begin{figure}	
	\includegraphics[width=8.5truecm]{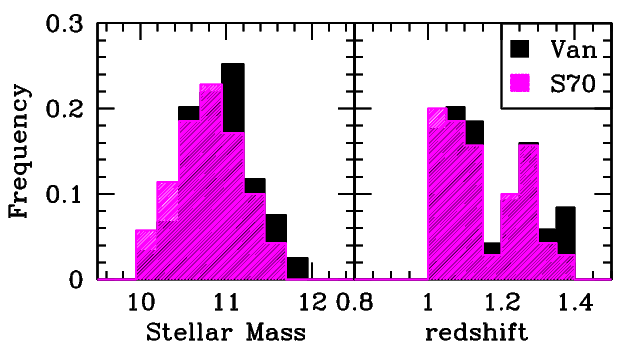}
   \caption{\label{fig:distrib} Stellar mass and redshift distributions of the 70 
   galaxies of our sample (shaded magenta histogram) are compared to the distributions of the parent
   VANDELS sample of passive galaxies (black histogram) in the redshift range 1.0$<$$z$$<$1.4.
   }
\end{figure}

The 70 passive galaxies studied here were extracted from the sample of 
268 UVJ-selected passive galaxies at $1.0<z_{phot}<2.5$ targeted by the VANDELS survey \cite[see][for a 
detailed description of the sample selection and spectroscopic observations]{mclure18,pentericci18} 
according to the following criteria. 
We first selected all the passive galaxies (119) with spectroscopic redshift in the 
range $1.0$$<z_{spec}$$<1.4$
and reliability redshift flag $\ge$3 (i.e. probability of $z_{spec}$ to be correct $>$80\%).
The redshift selection allows us to sample the rest-frame wavelength range [2600-4200]\AA\  
for all the selected galaxies, {  and  [2600-4350]\AA\ for those at $z_{spec}$$<1.3$)}. 
In this wavelength range, the main mid-UV indices 
(e.g., MgII($\lambda$2800), MgI($\lambda$2852), FeI($\lambda$3000)) and a 
number of optical
(e.g., CN3883, CaIIH\&K, D4000, H$\delta$, Ca4227, G-band) spectral features lie.
Finally, {  on the basis of previous experience \citep{saracco19}, we selected galaxies whose spectrum has a S/N$>6$ per \AA\ 
over the rest-frame range [3400-3600] \AA\ to assure an average accuracy on
spectral indices of $\sim$15\% and reliable stellar population properties, and 
no truncation (due to technical problems)} over the range 
[3350-4350] \AA\ to allow a reliable fitting of the spectrum.
After this cleaning we remained with a sample of 70 galaxies
in the mass range 10.0$<$log(M$_*/M_\odot$)$<$11.7.
Fig. \ref{fig:distrib} compares the stellar mass and the redshift distributions of the 70 passive galaxies (magenta shaded histogram) with the distributions of the parent VANDELS 
sample of passive galaxies in the same redshift range (black histogram).

We morphologically classified galaxies as early-type (ETG), and late-type (LTG, 
spiral S and irregular I) 
by inspecting the HST images for the 50 galaxies covered by HST observations.
The classification results into 18 LTGs (6 irregulars and 
12 spirals), 32 ETGs and 20 unclassified (for which we thus expect $\sim$13 ETGs 
and 7 LTGs).
Among the 18 LTGs, 6 galaxies were found to be a superposition or a merger of two galaxies.
These 6 galaxies were not included in the analysis (even if the fitting to their
spectra was performed, see below), resulting in a final sample of 64 passive galaxies,
of which $\sim$70 per cent are ETGs, as seen also in other samples covering similar
mass range
\citep[see e.g.,][]{tamburri14}.
{ We do not find a dependence of the fraction of LTGs on the mass of the galaxies}.
Hereafter, we refer to the sample of 64 galaxies as passive galaxies.

\subsection{Stellar population models}
\label{subsec:models}
In this analysis, we adopt as reference the UV-extended EMILES simple stellar 
population (SSP) models
\citep{vazdekis16, vazdekis15}, based on BaSTI isochrones \citep{pietrinferni04}
 and \cite{chabrier03} stellar initial mass function (IMF) \citep[see e.g.,][for a
 comparison among models and different IMFs in full spectral fitting]{ge19}.
We considered ages in the range [0.1; 5.0] Gyr, 
and metallicity [Z/H] in the range [-2.27; 0.26].\footnote{Models with [Z/H]=0.4,
although available, were not used in this analysis given the lower quality 
than the other models \citep[see][for details]{vazdekis15}}
These models have a FWHM spectral resolution of 3 \AA\ at $\lambda$$<$3540 \AA\ and
2.5 \AA\ in the optical domain \citep{vazdekis16}, 
higher than the rest-frame resolution ($\sim$6-7 \AA) of the VANDELS spectra.
We adopted, as reference, the set of EMILES ``base'' models for which it is assumed
that [Fe/H]=[Z/H], although this is only true for [Z/H]$\ge$0.0. 
For low metallicities the input stars are $\alpha$-enhanced.
Therefore, when a model with total metallicity [Z/H] is selected,  its [Fe/H] is lower \citep[see][for a detailed description]{vazdekis16}.

\section{Stellar age and metallicity estimates}
Stellar metallicities and ages for the whole sample of passive galaxies at $1.0<z<1.4$
were derived through non-parametric full-spectrum fitting (npFSF) performed over
the rest-frame wavelength range [3350-4350] \AA.
We limited the fit to 3350 \AA\ to exclude the UV regime where observations for local 
galaxies are not available (as this wavelength range is not accessible from the ground).
This allows us to perform a consistent comparison down to $z\sim0$ (see \S 4.2).
We used the \texttt{STARLIGHT} code \citep{cid05,cid07} and E-MILES models.
This code perform a fit by linearly combining SSPs with different ages Age$_i$ and 
metallicities Z$_i$, each one contributing with a different weight to the light and to 
the stellar mass.
{ 
Light-weighted ($L$) and mass-weighted ($M$) age Age$_{L,M}$ and metallicity [Z/H]$_{L,M}$ 
are thus defined
according to the relations \citep[e.g.,][]{asari07}
\begin{equation} 
{\rm Age}_{L,M}=\sum_i w_i(L,M) {\rm Age}_i
\end{equation}
 and
 \begin{equation}
{\rm [Z/H]}_{L,M}=log \sum_i w_i(L,M)Z_i/Z_\odot
\end{equation}
where $w_i$(L,M) are the light- and mass-weights.
}
An advantage of this fitting approach compared to the parametric FSF is that no 
a-priori assumption on the SFH is done, assumption which might 
significantly affect the resulting duration of the star-formation, 
as well as the inferred age and metallicity.

Before studying the relations among the stellar properties, we first assess their dependence 
on the stellar population models and on the main assumptions considered in the fitting, 
and the difference between luminosity-weighted and mass-weighted quantities.
This analysis (presented in Sections 3.1 ad 3.2) was performed on stacked spectra having 
higher signal-to-noise than individual spectra to minimize the uncertainties.


\subsection{Stacked spectra}
\begin{figure*}	
	\includegraphics[width=15.5truecm]{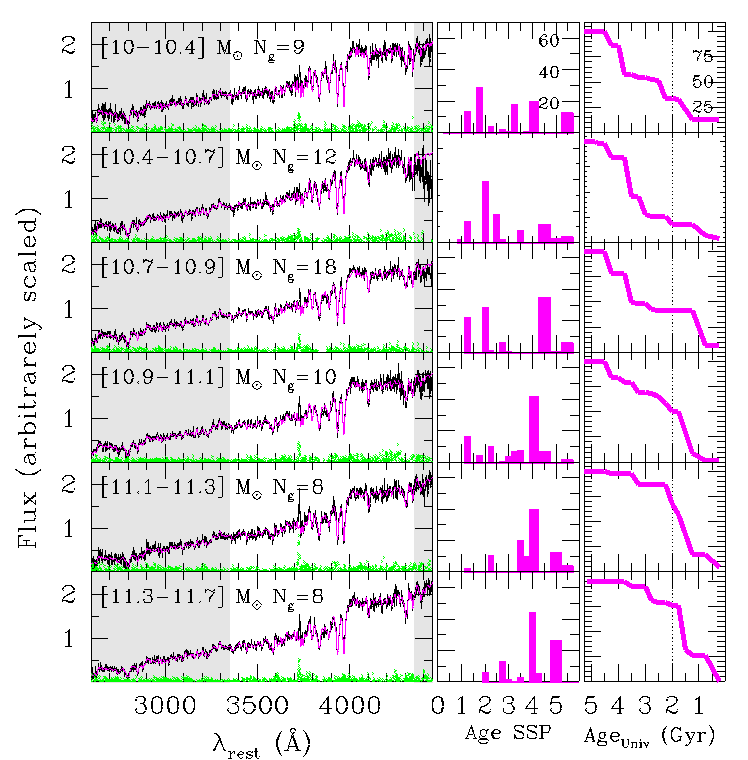}
   \caption{\label{fig:stack}  Left panels - Stacked spectra (black curves) of galaxies in six stellar mass ranges. Magenta curve is the best-fitting 
   composite model resulting from full-spectrum fitting performed with
   \texttt{STARLIGHT} over the wavelength range 3350-4350 \AA\ with EMILES 
   models (see \S 3). {  Residuals are shown in green.}
   In each panel, the mass range (in logarithmic form) and the number of stacked spectra
   are shown.
   The gray shaded regions mark the wavelength ranges masked in the fit:
   the UV wavelength range inaccessible 
   from the ground for galaxies at $z$$<$0.4 and $\lambda$$>$4350 \AA
   sampled only for few galaxies. 
 Central panels - Star formation history of galaxies. Fraction of stellar mass associated
 with each SSP contributing to the best-fitting composite model as a function of age.
 The numbers inside the upper panel mark the fractions 20\%, 40\% and 60\% respectively.
 Right panels - Cumulative SFH showing the growth of stellar mass as a function of the
 age of the cosmic time.
 { To go from the ages of the SSPs to the epoch of formation $t_{form}$ (see Sec. 5.1) 
 we considered $<$$z$$>$=1.2 (Age$_{Univ}$($z$=1.2)$\simeq$5.05 Gyr) as mean redshift of the stacks.} 
 The numbers inside the upper panel mark 25\%, 50\% and 75\% of the total mass, respectively.
 { The dotted line marks 2 Gyr ($z$$\simeq$3.0) just as reference.}
 }
 \end{figure*}
 
We divided the sample of 64  VANDELS passive galaxies in six mass ranges between 
10$^{10}$ M$_\odot$ and 10$^{11.7}$ M$_\odot$ to derive 
a single stacked spectrum representative of galaxies in each mass range.
Mass ranges were chosen to have a minimum of 8 galaxies in each of them.
For the stacking, each spectrum was first shifted to the rest-frame, then normalized
to the mean flux measured in the rest-frame wavelength range 3350-3550 \AA, flat and
free from significant features, and finally re-sampled to a common dispersion of 1 \AA/pix. 
The spectra were then median stacked and the uncertainties were calculated 
using the Median Absolute Deviation (MAD) estimator.
Fig. \ref{fig:stack} shows the stacked spectra corresponding to the six
mass bins.
The resulting signal-to-noise ratios are in the range SNR$\simeq18-25$ \AA$^{-1}$.

{ The stacked spectra of galaxies with mass log(M$_*$/M$_\odot$)$>$11.0 show
weak [OII]($\lambda$3727) emission lines, not detectable in the individual spectra
\citep[see also][for OII emission in LEGA-C passive galaxies]{maseda21}.
The measured flux\footnote{Flux was estimated by fitting a Gaussian function to the 
line after having removed the underlying continuum evaluated through a polynomial 
fitting of the regions adjacent to the line.}
of the strongest [OII] emission is F(OII)=6.1$\pm0.6$$\times$10$^{-18}$ 
erg cm$^{-2}$ s$^{-1}$ associated to the stack of galaxies in the 
mass range [11.1-11.3] M$_\odot$.
This flux corresponds to a luminosity L(OII)=5.0$\pm0.5$$\times$10$^{40}$
erg s$^{-1}$ at the mean redshift $z$=1.2.
Assuming that this emission is due only to SF, and that the local relation 
SFR(M$_\odot$ yr$^{-1}$)=(1.4$\pm$0.4)$\times$10$^{-41}$L(OII) 
\citep{kennicutt98} is valid also at higher redshift, we derive a 
SFR=0.7$\pm$0.3 M$_\odot$ yr$^{-1}$.
Therefore, the current residual star formation for
galaxies with log(M$_*$/M$_\odot$)$>$11.0 of our sample is, on average,
lower than SFR$\sim$1.0 M$_\odot$ yr$^{-1}$ and decreases towards lower masses.
{  We note that no differences are obtained in the full spectral fitting 
by masking or not the spectral regions with the above emission lines. }

\subsection{Dependence on models, fitting and definitions \label{sec:dependence}}
We performed non-parametric FSF to the stacked spectra shown 
in the left panels of Fig. \ref{fig:stack} using three different sets of models and 
two fitting codes 
to asses and quantify the dependence of the results on these basic fitting assumptions.
{ We underline the fact that the following analysis does not want to be an exhaustive comparison between the SSP models in the literature nor the spectral fitting codes, but is simply aimed at verifying if the results can depend on their choice.}
Besides the EMILES models, we considered the 2016 updated version of the \cite{bruzual03} models (CB16 hereafter) with a Chabrier IMF, ages in the range [0.1; 5.0] Gyr and metallicity in the range [-2.3; 0.4], and the \cite{maraston11} models (M11 hereafter) with Chabrier IMF, ages in the range 
[0.1; 5.0] Gyr and metallicity in the range [-1.3; 0.3].
These three sets of models are based on the same MILES 
spectral stellar library in the optical \AA\ \citep[i.e. $\lambda$$>$3500\AA; ][]{falcon11}.
{ The exclusion of the UV spectral range ($\lambda$$<$3350 \AA) from the fit 
allows us, among other things, a fair comparison between these different SSP models.}

As an alternative to the \texttt{STARLIGHT} code, we also considered the \texttt{pPXF} code \citep{cappellari17}.
Both codes perform the non-parametric FSF by linearly combining SSPs extracted from
the same base of spectral templates.
The best fitting composite model is found by $\chi^2$ minimization.
The main differences between the two codes are that \texttt{pPXF} performs a regularization 
of the solutions \citep[see][]{cappellari17} allowing to explore their degeneracy 
while \texttt{STARLIGHT} does not, and that this latter explores the composite models 
using a Markov Chain Monte Carlo algorithm \citep[see][for a comparison of the performances of the two codes]{ge18}.
\begin{figure}	
 	\includegraphics[width=8.truecm]{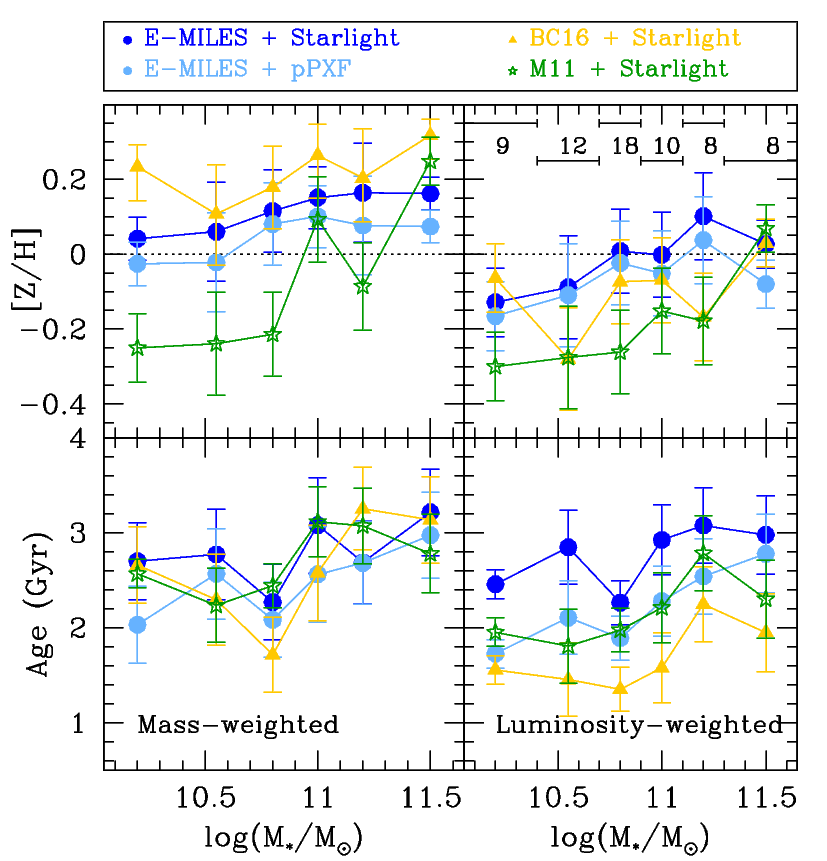}
   \caption{\label{fig:comparison} Metallicity (top) and age (bottom) estimates as a
   function of stellar mass. Left and right panels show mass- and luminosity-weighted 
   quantities respectively, resulting
   from full-spectrum fitting of stacked spectra of passive galaxies in different mass ranges (Fig. \ref{fig:stack}) for different stellar population models and fitting codes.  
   Blue and light-blue filled circles are the results obtained with EMILES SSPs using \texttt{STARLIGHT} and \texttt{pPXF} fitting code respectively; yellow triangles and green stars represent STARLIGHT fitting with CB16 and M11 models respectively { (see note 
   \ref{foo:offset} for the $\pm$0.06dex offset between models)}. 
   The number of galaxies contributing to the stack in each bin is shown on top of right panel together with the width of the bins.
  }
\end{figure}

\begin{figure}	
	\includegraphics[width=8.5truecm]{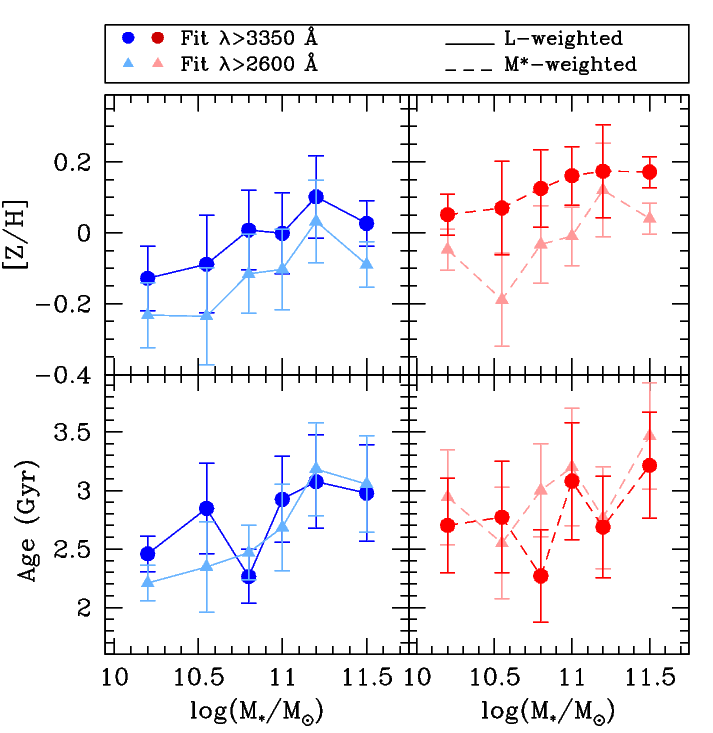}
   \caption{\label{fig:comparisonUV} Stellar age (lower panels) and metallicity
   (upper panels) resulting from full-spectrum fitting of stacked spectra (Fig. \ref{fig:stack}) in different 
   wavelength ranges are shown as a function of stellar mass for EMILES SSP models.  
   Filled circles and triangles are values resulting from fitting the spectra at 
   $\lambda_{rest}>3350$ \AA\ and $\lambda_{rest}>2600$ \AA, respectively.
   Solid and dashed lines marks luminosity-weighted and mass-weighted values respectively.
  }
\end{figure}

In Fig. \ref{fig:comparison} the stellar metallicity and age resulting from the fitting
of the stacked spectra are shown as a function of stellar mass for the two fitting codes 
and the different models. 
For E-MILES models, the two fitting codes provide consistent values 
both for metallicity and ages.
{ We notice, however, that luminosity-weighted ages from \texttt{pPXF} are systematically 
younger than those from \texttt{STARLIGHT}}.
The resulting values are summarized in Tab. \ref{tab:parameters}.

On the contrary, significant systematics (of the order of $\sim$0.2-0.4dex) 
are present among the values obtained with different models.
In particular,  
the metallicity [Z/H] (Fig. \ref{fig:comparison}, upper panel) 
obtained for M11 models is sistematically lower 
than the metallicity obtained with EMILES and CB16 models,
being on average, [Z/H]$_{L,EMILES}=$[Z/H]$_{L,M11}$+0.17$\pm$0.08
for the luminosity-weighted values, { offset that reduces
to +0.11$\pm$0.08 considering the different solar metallicity of models}\footnote{\label{foo:offset}Note that solar
metallicity of EMILES models based on BASTI isochrones \citep{pietrinferni04} differs by
+0.06dex with respect to models based on PADOVA isochrones \citep{girardi00}, such as
CB16 and M11 \citep[see ][]{vazdekis15}.}. 
The discrepancy between CB16 and M11 is even higher 
($\Delta[Z/H]$$\sim$0.3) when mass-weighted metallicity is considered.
For the age, the differences among the mass-weighted values 
obtained with the different models are not significant.
However, a significant systematic is found for CB16 models when the 
luminosity-weighted stellar age is considered, being on average 
Age$_{L,EMILES}$=1.5($\pm$0.08)$\times$Age$_{L, CB16}$.
Therefore, except for the mass-weighted age, which all models return consistent values of, luminosity-weighted age and metallicity values depends on the models assumed in the analysis.

\begin{table*}
\begin{minipage}[t]{1\textwidth}
\caption{\label{tab:parameters} Mean stellar age and metallicity of galaxies in different 
ranges of stellar mass as resulting from the full-spectrum fitting$^a$ of the stacked 
spectra perfomed with \texttt{STARLIGHT} (S) and \texttt{pPXF} (p).}
\centerline{
\begin{tabular}{clrrrcl}
\hline
\hline
Log(M$_*$)   &   Age$_L$& [Z/H]$_L$& Age$_{M*}$& [Z/H]$_{M*}$& A$_V^b$& Code\\
(M$_\odot$)  &  (Gyr)   &          & (Gyr)     &             & (mag)  & \\
\hline
10.20 &  2.4$\pm$  0.2 & -0.13$\pm$ 0.09 &  2.7$\pm$ 0.4 &  0.05$\pm$  0.06 & 0.00& S \\
''    &  1.7$\pm$  0.2 & -0.17$\pm$ 0.09 &  2.0$\pm$ 0.4 & -0.04$\pm$  0.06 & ....& p \\
10.55 &  2.8$\pm$  0.4 & -0.09$\pm$ 0.10  &  2.7$\pm$ 0.5 &  0.07$\pm$  0.10  & 0.16& S \\
''    &  2.1$\pm$  0.4 & -0.11$\pm$ 0.10  &  2.6$\pm$ 0.5 & -0.03$\pm$  0.10 & .... & p \\
10.80 &  2.3$\pm$  0.2 &  0.01$\pm$ 0.10  &  2.3$\pm$ 0.4 &  0.12$\pm$  0.10 & 0.00 & S \\
''    &  1.9$\pm$  0.2 & -0.02$\pm$ 0.10  &  2.1$\pm$ 0.4 &  0.07$\pm$  0.10 & .... & p \\
11.00 &  2.9$\pm$  0.4 &  0.00$\pm$ 0.10  &  3.1$\pm$ 0.5 &  0.16$\pm$  0.08 & 0.00& S \\
''    &  2.3$\pm$  0.4 & -0.05$\pm$ 0.10  &  2.6$\pm$ 0.5 &  0.09$\pm$  0.08 & ....& p \\
11.20 &  3.1$\pm$  0.4 &  0.10$\pm$ 0.10  &  2.7$\pm$ 0.4 &  0.17$\pm$  0.10 &  0.24& S \\
''    &  2.5$\pm$  0.4 &  0.04$\pm$ 0.10  &  2.7$\pm$ 0.4 &  0.07$\pm$  0.10 &  ....& p \\
11.50 &  3.0$\pm$  0.4 &  0.03$\pm$ 0.06 &  3.2$\pm$ 0.5 &  0.17$\pm$  0.04 & 0.33& S \\
''    &  2.8$\pm$  0.4 & -0.08$\pm$ 0.06 &  3.0$\pm$ 0.5 &  0.06$\pm$  0.04 & ....& p \\
\hline
\end{tabular}
}
{$^a$ The fitting was performed with EMILES models. Luminosity-weighted and mass-weighted values are marked with $_L$ and $_{M*}$ respectively.
$^b$ Internal reddening for different extinction laws is allowed by 
\texttt{STARLIGHT}, while \texttt{pPXF} uses multiplicative polinomials
to correct low frequency continuum variations.}
\end{minipage}
\end{table*}

A well known, but important systematic difference exists between luminosity- and mass-weighted
values independently of the models considered, as shown in  Fig. \ref{fig:comparison} \citep[see also][]{barone20}.
As for the metallicity, the difference is particularly significant being,
on average,
[Z/H]$_{M*}\simeq$[Z/H]$_L$+0.2($\pm$0.05), while it is less important for
the age,
Age$_{M*}$=1.05($\pm$0.08)$\times$Age$_L$.
The systematic difference is due to the different M/L values of SSPs seen at
different age and metallicity.
Therefore, as expected, considering luminosity-weighted or mass-weighted values implies
significant systematic differences.

Finally, the spectral range considered in the fitting can affect the
resulting stellar population properties.
This is shown in Fig. \ref{fig:comparisonUV} where full-spectrum fitting
of stacked spectra was performed with \texttt{STARLIGHT} and EMILES models over two 
wavelength ranges, 2600 \AA$<$$\lambda_{rest}$$<$4350 \AA\ and 
3350\AA$<$$\lambda_{rest}$$<$4350 \AA.
Also in this case, the largest systematic is seen for the metallicity.
The inclusion in the fitting of the UV wavelength range 2600-3350 \AA, typically
missed for ground-based observations of galaxies at $z<0.4$, 
results in metallicities systematically lower 
(by $\sim$0.15dex) then those obtained by fitting the range $\lambda_{rest}>3350$ \AA.
This is true both for luminosity-weighted and mass-weighted values.
{  It is worth noting that this systematic is not dependent on the models used.
Indeed, we obtained the same result with BC16 models, 
which extend to UV:
by fitting the spectral range [2600-4350] \AA, we derived metallicities, on average,
$\sim$0.15dex lower then fitting the range [3350-4350] \AA.}

These metallicity offsets could be due to the combined effect of the different stellar
populations sampled by the two different wavelength ranges, and by the poorer knowledge 
and implementation of the UV properties in the stellar population synthesis models \citep[see,  e.g.,][]{maraston09,vazdekis16,lecras16,lonoce20}.
{  Simulations cannot help in disentangling these effects.\footnote{  By simulating a spectrum with SSP models and fit it with SSP models, we cannot test how the poor knowledge of the UV spectral range may affect the results since different spectral regions would be, by construction, self-consistent.
However, in Appendix \ref{sec:appsimul}, we simulate a galaxy with 
mixed stellar population to show qualitatively how the inclusion in the fit 
of the UV spectral range may affect the metallicity estimate.}} 
It is important to note that, since the observed spectral range at rest shifts toward shorter wavelengths with increasing redshift, the effect above would result in lowering the metallicity values at high redshift with respect to those at lower redshift, mimicking an evolution.

Any comparison of the stellar population properties of different samples of galaxies
cannot neglect the dependencies and systematics derived above.
This aspect becomes decisive in detecting evolution of these properties across cosmic time 
through the comparison of measurements at different redshift.
The analysis shows that such kind of measurements must be homogeneous to be comparable
and that the variation (if any) should be considered relative since 
values cannot be considered absolute, being dependent on models, 
methodology and spectral range.

In the next sections, we first determine the main relations among stellar population 
properties of passive galaxies at $z\sim1.2$, then we probe their evolution with redshift 
using homogeneous measurements and method for different galaxy samples.

\section{The stellar mass-metallicity relation} 
\label{sec:metal}
\subsection{The MZR of passive galaxies at $z\sim1.2$}
Fig. \ref{fig:comparison}
shows that the stellar metallicity of passive galaxies at $z\sim1.2$ increases with 
stellar mass independently of the models assumed in the fitting and of the 
values considered, i.e., luminosity- or mass-weighted.
This is also confirmed by the best fitting linear relations obtained for the different 
models reported in Tab. \ref{tab:relations}.
The best fitting parameters reflect the large differences among the 
spectral fitting obtained for different models, 
with the large variation of the zeropoint $b$ showing the large systematic 
in the [Z/H] values discussed in the previous section.

\begin{table}
\caption{\label{tab:relations} Best fitting linear relations 
to the luminosity- and mass-weighted  metallicity values derived from \texttt{STARLIGHT}
FSF to the stacked spectra obtained in the different mass ranges, 
for different models
[Z/H]=a$_{L,M}$log(M$_{11}$)+b$_{L,M}$, where M$_{11}$=M$_*$/(10$^{11}$M$_\odot$).
}
\centerline{
\begin{tabular}{lllll}
\hline \hline
a$_L$      &       b$_L$ &	    a$_{M}$	  &	  b$_{M}$   & Models \\ 
\hline    
0.15 $\pm$ 0.06 &  0.01 $\pm$0.02   & 0.11 $\pm$ 0.06&  0.14 $\pm$  0.01 & EMILES \\
0.09 $\pm$ 0.09 & -0.09 $\pm$0.05   & 0.09 $\pm$ 0.06&  0.23 $\pm$  0.03 & CB16   \\
0.26 $\pm$ 0.07 & -0.15 $\pm$0.03   & 0.39 $\pm$ 0.10&  -0.07 $\pm$ 0.05 &  M11 \\
\hline
\end{tabular}
}
\end{table}

The stellar MZR of passive galaxies 
at $z\sim1.2$ in the mass range  10$\leq log(M_*/M_\odot)<11.6$ 
resulting from npFSF with \texttt{STARLIGHT} and E-MILES models 
{  to stacked spectra are}
\begin{equation}
\label{eq:mz}
 \begin{split}
 [Z/H]_L=0.15\pm0.06 \ log(M_{11})+0.01\pm0.02 \\ 
 [Z/H]_{M*}=0.11\pm0.06 \ log(M_{11})+0.14\pm0.01 
 \end{split} 
 \end{equation}
for luminosity-weighted and mass-weighted values, respectively (first line of Tab. \ref{tab:relations}), where M$_{11}$=M$_*$/(10$^{11}$M$_\odot$).
\begin{figure*}	
\includegraphics[width=12.5truecm]{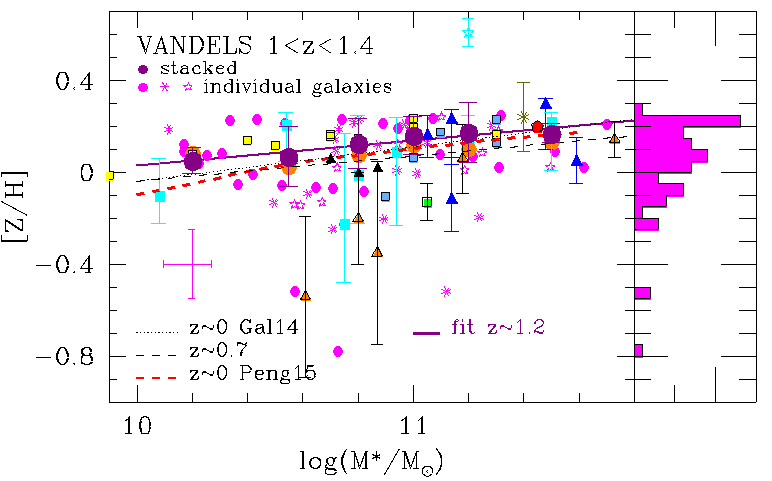}
\includegraphics[width=4.5truecm]{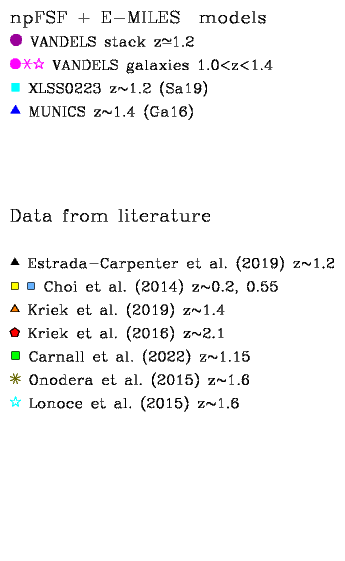}	
   \caption{\label{fig:mass_Z} 
   Mass-weighted metallicity derived through 
   non-parametric full-spectrum fitting (npFSF) with \texttt{STARLIGHT} and E-MILES models
   as a function of mass for quiescent galaxies at comparable redshift $z$$\sim$1.2. 
   Magenta symbols (filled circles, stars and skeletals) are VANDELS ETGs, LTGs 
   and unclassified galaxies, respectively, at 1.0$<$$z$$<$1.4;
   {  big orange filled circles are the median values of VANDELS galaxies in the same mass bins of stacked spectra;}
   big purple filled circles are VANDELS stacked spectra (errorbars are the standard deviations of the values in each bin);
   cyan filled squares are individual passive galaxies in cluster XLSS0223 at $z\sim1.2$ from
   \citet[][Sa19]{saracco19}; blue triangles are individual massive ETGs at $z\sim1.4$ from \citet[][Ga16]{gargiulo16}.
   The purple continuous line is the best fitting linear relation to the stacked values 
   of VANDELS (see Eq. \ref{eq:mz}).
   The black dotted and dashed lines are the MZRs
   of quiescent galaxies at $z$=0.1 and $z$=0.7, respectively, from \citet[][Gal14]{gallazzi14}. The red dashed line is the observed stellar MZR of local passive galaxies derived by \citet{peng15} from SDSS data.
Data from the literature, as derived by \citet{estrada19},
   \citet{choi14}, \citet{kriek19}, \citet{kriek16}, \citet{carnall22}, \citet{onodera15} and
   \citet{lonoce15}, are superimposed to npFSF+EMILES values. 
   Symbols are as in the legend.  
   Cross in the lower left of the left panel represents the typical error
   at 1$\sigma$ of individual measurements for VANDELS galaxies. 
   }
\end{figure*}

Fig. \ref{fig:mass_Z} shows the mass-weighted metallicity as a function of mass for the 
individual VANDELS passive galaxies (magenta symbols)\footnote{  The typical uncertainty on single measurement has been derived according to the 
following procedure. We considered the best fit composite model of one of the VANDELS spectra 
with an average S/N$\sim$7-8 \AA$^{-1}$, representative of the selected spectra.
We obtained a number of realizations by summing to this template the shuffled
residuals of the fit itself. For each realization, we perform the fit 
and estimate the metallicity. As typical error, we 
considered the standard deviation from the realizations.} superimposed to the stacked values 
(purple filled circles).
The purple solid-line is the mass-weighted relation reported in Eq. \ref{eq:mz}.
The npFSF was performed with \texttt{STARLIGHT} and E-MILES SSPs over the wavelength range 
$\lambda_{rest}>3350$ \AA.
Internal reddening in the range A$_V$=0-2 mag was allowed in the fitting by considering  
the \citet{calzetti00} extinction law.
In Fig. \ref{fig:av} the distributions of the values of extinction and of the $\chi^2$ 
   values resulting from the fitting are shown.
\begin{figure}	
	\includegraphics[width=8.0truecm]{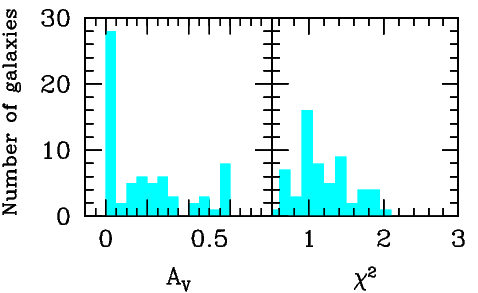}
   \caption{\label{fig:av} Distributions of extinction values (left) and of $\chi^2$ 
   values (right) resulting from the npFSF of VANDELS galaxies
   with \texttt{STARLIGHT} and E-MILES models}
\end{figure}

The stellar metallicity of $\sim$95 per cent of the sample falls in the range 
-0.35$<$[Z/H]$<$0.25, 
as shown by the distribution in the right panel of Fig. \ref{fig:mass_Z}.
In particular, 67 (58) per cent of the galaxies at $z\sim1.2$ have mass-weighted 
(luminosity-weighted) metallicity higher than solar ([Z/H]$>0$).
In Fig. \ref{fig:mass_Z} we also show, for comparison, the metallicity of massive 
(log(M$_*$/M$_\odot$)$>$11) ETGs at $z\sim1.4$ derived with the same procedure adopted 
in this work from the VLT-FORS2 spectra studied by \citet[][Ga16; blue triangles]{gargiulo16}, 
and for passive galaxies in cluster XLSS0223 at $z\sim1.2$ from LBT-MODS spectra
\cite[][Sa19; cyan filled squares]{saracco19}.
The agreement among the different data at comparable redshift is very good.

We confirm that the stellar metallicity of passive galaxies is positively correlated 
with their stellar mass at $z\sim1.2$, as tentatively previously found by \cite{kriek19} and
\cite{saracco19} on much smaller samples.
This trend is independent of the models assumed\footnote{  The Spearman rank test
performed on the metallicity values derived from the stacked spectra for the different models and codes (Fig. \ref{fig:comparison}) provided correlation coefficients $\rho_s$$>$0.92 (i.e., probabilities $p_s$$\sim$0.005 that the data 
are not correlated) in all the cases, with the exception of CB16 that provided 
$\rho_s$$\simeq$0.54 and $p_s$$\simeq$0.26. 
The test performed on single VANDELS values provided $\rho_s$$\simeq$0.27 with an
associated probability $p_s$$\simeq$0.04.}, as shown in Fig. \ref{fig:comparison}, even if different models provide
systematic differences in the derived stellar populations properties.
The scatter in the metallicity of individual VANDELS galaxies at 
log(M$_*$/M$_\odot$)$>$11 is about $\pm$0.1dex, while it is larger at 
lower masses, where galaxies span a wider range of metallicity.
It is worth noting that the positive trend between metallicity and mass is mainly
due to the lack of low-metallicity galaxies with high mass rather than to a higher metallicity
of high-mass galaxies, as also seen in the local Universe  \cite[e.g.,][]{gallazzi05,asari09, choi14, mcdermid15}, and this produces also the flattening of the relation for masses 
log(M$_*$/M$_\odot$)$>$11.2.

\subsection{The evolution of the stellar MZR}
\begin{figure*}	
\includegraphics[width=12.5truecm]{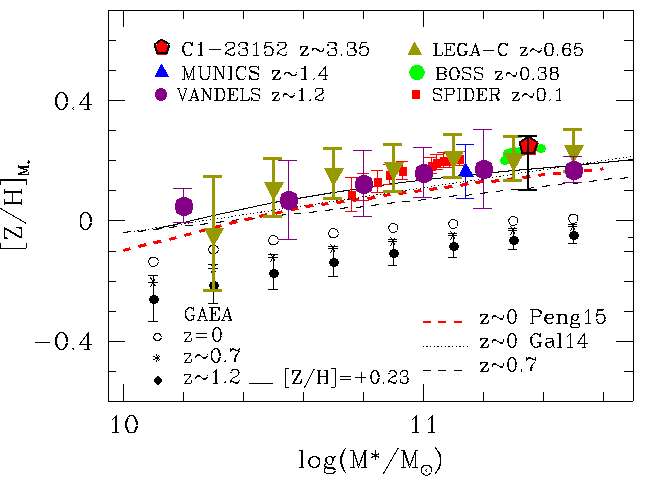}
  \caption{\label{fig:mass_Zevol} 
   Mass-weighted metallicity derived through 
   non-parametric full-spectrum fitting (npFSF) to stacked spectra of quiescent galaxies 
   at different redshifts as a function of mass. 
   The fitting was performed with \texttt{STARLIGHT} code and
   EMILES models over the same wavelength range ([3350-4350]\AA) for all the spectra.
  Purple circles are VANDELS stacked spectra, military triangles 
   are LEGA-C stacked spectra at 0.6$<$$z$$<$0.7; green filled circles are 
  stacked BOSS spectra of massive ETGs at $\langle z\rangle\sim$0.38 from \citet{salvador20};
   red squares are stacked SPIDER spectra of massive ETGs at $z\sim0.05$ \citep{labarbera13};
    blue triangle is the median value of the 5 massive ETGs at $z\sim1.4$ from \citet{gargiulo16}; red diamond is the massive ETG C1-23152 at $z$$\sim$3.35 studied by
   \citet{saracco20apj}.
   Superimposed to the data are also predictions from GAEA models \citep[][black open
   and filled symbols]{fontanot21}
   for different redshifts as in the legend.
   { The black thin curve is the GAEA prediction at $z$$\sim$1.2 rescaled up
   by +0.23 in [Z/H] to match the median value of VANDELS data.}
   The mass-metallicity relations by \citet[][black dotted and dashed lines]{gallazzi14}
   and \citet[][red dashed curve]{peng15} are also shown.
   }
\end{figure*}

In Fig. \ref{fig:mass_Zevol} we compare the stellar metallicity of 
passive galaxies over the redshift range 0$<$$z$$<3.35$ derived homogeneously 
according to the procedure adopted in this work, i.e., same fitting code (\texttt{STARLIGHT}), simple stellar population models (EMILES) and fitting wavelength range ([3350-4350]\AA).
At redshifts lower than VANDELS data, $z$$<$1.2, we derived stellar metallicity
from stacked spectra (S/N$>15$ \AA$^{-1}$) of passive galaxies selected from the LEGA-C survey 
\citep{vanderwel16,vanderwel21} in the mass range 10.2$<$log(M$_*$/M$_\odot$)$<$11.6 
and in the redshift range 0.6$<$$z$$<$0.7 (see Bevacqua et al, in preparation).
The metallicity values at $z$$\sim$0.65 derived at the different 
masses agree with those at $z$$\sim1.2$ derived from VANDELS stacked spectra. 
Also the LEGA-C data show a positive correlation between metallicity and mass.

At $z$$\sim$0.38 we derived the metallicity of massive ETGs
from the high S/N ($\sim$100) stacked spectra 
of \citet{salvador20}.
They selected ETGs in the narrow mass range 11.2$<$log(M$_*$/M$_\odot$)$<$11.45 from BOSS and
stacked the spectra according to their velocity dispersion 
(220$<$$\sigma_v$$<$340 km s$^{-1}$) in bin of 100 km s$^{-1}$
considering, for each stack, the median stellar mass
\citep[see][for details]{salvador20}.
The metallicity values we derived lie on the relation described by 
the LEGA-C data  
and are consistent with those from VANDELS data at $z$$\sim$1.2.

At even lower redshift, $z$$\sim$0.05, we used the high S/N ($\sim$100) stacked spectra 
of ETGs selected from the SPIDER sample \citep{labarbera13} in the mass range 
10.6$<$log(M$_*$/M$_\odot$)$<$11.2.
The stacking was made according to their velocity dispersion as in \citet{salvador20}.
Also in this case, the metallicity values agree with those at higher redshift.

At redshift higher than VANDELS data, we considered the mean metallicity value (blue triangle) of the 5 massive (11$<$log(M$_*$/M$_\odot$)$<$11.6) ETGs at $z$$\sim1.4$ studied by \citet{gargiulo16} and shown individually in Fig. \ref{fig:mass_Z}, and the metallicity
derived for the massive 
(log(M$_*$/M$_\odot$)$\sim$11.3) ETG C1-23152 at $z$$\sim$3.35, whose estimate was obtained 
following the same procedure adopted in this work \citep{saracco20apj}. 
The metallicity of these galaxies is consistent with the values derived for VANDELS galaxies
and for massive ETGs at $\sim1.4$ at similar mass, as well as with those derived at lower redshift for LEGA-C, BOSS and SDSS massive ETGs.

Therefore, our analysis does not show evidence of metallicity evolution
in the redshift range probed.
In particular, we do not detect change  of the 
stellar metallicity of passive galaxies with mass log(M$_*$/M$_\odot$)$>$11
at least in the redshift range 0.1$<$$z$$<$1.4.
{ At higher redshift, we note that the metallicity of C1-23152, the only
massive passive galaxy for which an estimate of the stellar metallicity 
has been obtained to date at $z$$>3.0$ suggests a lack of 
variation even up to these redshifts.}
In fact, the metallicity values derived from the high S/N stacked spectra in the above 
mass range, do not show any trend with redshift, and the metallicity values of 
individual massive galaxies are consistent with each other up to the highest redshift probed 
by our data.
The stellar MZR defined over the mass range  
10$<$log(M$_*$/M$_\odot$)$<$11.6 does not show evolution
between $z$$\sim$1.2 and $z$$\sim$0.65, either in the mean value or in the slope.

It is worth noting that some hydrodynamic simulations and semi-analytic models predict an evolution of the stellar metallicity of galaxies with redshift, even if typically less 
than 0.1 dex \citep[see, e.g., Fig. 11 in][]{guo16}.
However, the direction of the metallicity evolution (increasing or decreasing with redshift) 
is not always coincident among them.
In Fig. \ref{fig:mass_Zevol} the mass-metallicity for passive galaxies as 
resulting from the predictions of the GAlaxy Evolution and Assembly (GAEA) 
semi-analytic model \cite[][]{fontanot21} is also shown as an example.
Models show the relation for mass-weighted values expected at $z$$\sim$1.2, 
$z$$\sim$0.7 and $z$$\sim$0 
for passive galaxies defined according to their specific star formation rate log(sSFR)$<$-0.15$\times$log(M$_*$)-9.66 \citep{gallazzi21}\footnote{We note that the use of different criteria
to select passive galaxies (e.g., sSFR$<$0.3/t$_{Hubble}$($z$)) does not introduce significant differences in the result.}.
The agreement between the slope of the predicted {  (0.14$\pm$0.05)} and observed
relation (see eq. \ref{eq:mz}) is remarkable,
as well as the flattening of the relation at large masses, { as shown in Fig. \ref{fig:mass_Zevol} where the GAEA predicted MZR at $z$$\sim$1.2 is rescaled up 
by [Z/H]=+0.23 to match the median [Z/H] of VANDELS stacks.}
The predicted systematic increase of the metallicity with decreasing redshift 
is consistent with no evolution especially at log(M$_*$/M$_\odot$)$>$11, where
the evolution is much smaller than the scatter in the data.
{ We note that the possible tension between the normalization of the observed and predicted relations could be not significant.
Indeed, as shown in Sec. 3.2, different SSP models can provide metallicity values
differing even by 0.3dex (see Figure \ref{fig:comparison}), a difference that alone
could justify the apparent discrepancy.
Therefore, we believe that there are not the conditions to support a discrepancy between 
the GAEA predicted stellar metallicities and those observed.
We believe that the low mass regime of the relation, still difficult to 
probe with the current ground-based observing facilities, would deserve to be 
investigated since, at low masses, a larger evolution is predicted.
}

\subsection{Comparison with previous estimates of metallicity in the literature}
For completeness, we also report metallicity estimates from the literature, even if based 
on different models and methods.
Whenever possible, we distinguish mass-weighted from luminosity-weighted estimates given the 
 difference existing between these two quantities.
{ We remind the reader that a systematic difference of 0.06dex in metallicity exists
between EMILES and some other models (see note \ref{foo:offset}).
When relevant for the comparison, we will explicitly take this offset into account.
}

At low redshift, we show in Fig. \ref{fig:mass_Z} the SSP-equivalent metallicity 
derived by \citet{choi14}\footnote{\citet{choi14} derive the abundances of metal elements.
We obtained the metallicity using the relation [Z/H]=[Fe/H]+0.94[Mg/Fe] \citep{thomas03}.}
from stacked spectra of passive galaxies selected from the AGES 
survey \citep{kochanek12}.
We considered their estimates in the redshift range 0.1$<$$z$$<$0.3 and 0.4$<$$z$$<$0.7 
($z$$\sim$0.2 and $z$$\sim$0.55 in the legend, respectively). 
They used the fitting code and models developed by \cite{conroy12}.
Their data follow the MZR found by \citet{gallazzi06}
for local ($z$$\sim$0.1) ETGs (dotted line) and agree well with the
values we derived both at $z$$\sim$1.2 from VANDELS data and at $z$$\sim$0.65 from
LEGA-C data even when an offset by 0.06dex is considered. 

Besides these data, we show the metallicity value obtained by \cite[][; C22 sample hereafter]{carnall22} for stacked 
spectra of 91 massive passive galaxies in 
the redshift interval 1.0$<$$z$$<$1.3.
{They selected galaxies brighter than J=21.5 from the VANDELS photometric sample \citep{mclure18} applying a slightly different UVJ color selection
criterion to define passive galaxies with respect to VANDELS, and 
recomputed stellar masses.
The resulting sample is complete down to log(M$^*$/M$_\odot$)=10.8 and
composed of 77 galaxies with VANDELS spectra and 14 galaxies with KMOS spectra 
\citep[see][]{carnall22}.
We expect no more than 30 galaxies in common between our and C22 sample\footnote{ The actual overlap between our sample and C22 sample, as well as the overlap with the \cite{carnall19} sample, cannot be quantified as both samples are not disclosed.}, i.e., the galaxies
more massive than
log(M$_*$/M$_\odot$)$=$10.8 in our sample.}
Their metallicity estimate is based on parametric FSF performed with \texttt{BAGPIPES} code \citep{carnall18}
and CB16 models \citep{bruzual03} over the wavelength range [3550-6400] \AA\
thanks to KMOS observations.
The mean mass-weighted metallicity they obtained, [Z/H]=-0.13$\pm$0.08
{ ([Z/H]=-0.07 when corrected by +0.06dex),}
is $\sim$2$\sigma$ lower than our 
estimate for the same stellar mass (see Tab. \ref{tab:parameters}).
According to the comparison among models shown in Fig. \ref{fig:comparison}, the offset
between the metallicity values found in our and in their work cannot be due 
to the different models used (E-MILES vs CB16):
{ by adopting CB16 models we obtain a metallicity even higher than
the one obtained with E-MILES models, as shown in the upper-left panel of Fig. \ref{fig:comparison}.}

A possible reason for the discrepancy could be the different wavelength range 
considered in our and in their fitting procedure.
Indeed,  the range [4400-6400] \AA\ 
is not present in our spectra because of the lack of near-IR observations.
We verified whether the inclusion of this range of wavelengths in the fitting 
affects the metallicity estimate with respect to the estimate 
based on the range [3350-4350] \AA. 
By extending the fitting of high S/N BOSS stacked spectra of massive 
ETGs at $z$$\sim$0.38 \citep[][]{salvador20} up to $\lambda$$\sim$6000 \AA,  
we did not detect significant offset in the metallicity (see Appendix B, Fig. \ref{fig:comp_5850}).

{A more direct comparison can be made with the results obtained by \cite{carnall19},
{  based on the same spectra used in our analysis}.
In this case they selected 75 passive galaxies from the VANDELS spectroscopic sample in
the redshift range 1.0$<$$z$$<$1.3 at log(M$^*$/M$_\odot$)$>$10.3.
Considering that there are 5 galaxies in our sample with $z$$>$1.3 and 5 galaxies with
log(M$^*$/M$_\odot$)$<$10.3, we expect a sample of about 54 galaxies in common (60 galaxies 
considering also the 6 galaxies we removed being superposition or merger of two galaxies, see Sec. 2.1).}
A systematic difference between our and their [Z/H] estimates exists among the 
metallicity values for individual passive galaxies that they show in Fig. 3 of \cite{carnall22}:
none of the passive galaxies of their sample has metallicity higher than solar, contrary to a 
fraction of $\sim$60 per cent with supersolar metallicity in our sample (see above).
The spectral fitting in \cite{carnall19} was performed in the wavelength range [2600-4400] \AA.
We have already shown that the inclusion of the UV part of the spectrum in the fitting, 
not observed in local and low redshift ($z$$<$0.4) galaxies, 
leads to metallicity values systematically lower by 0.15dex (see \S\ 3.2 ).
However, from Fig. \ref{fig:comparisonUV}, it can be seen that the inclusion of the UV
wavelength range in our fitting, leads to a mean metallicity [Z/H]$\ge$0 for galaxies more
massive than log(M$_*$/M$_\odot$)$>$11.0, hence higher than the estimates in 
\citet[][]{carnall19}.
Moreover, as noticed above, by adopting CB16 models we would obtain metallicities 
even higher than those obtained with E-MILES models.
Finally, they do not find a positive trend of the metallicity with the stellar
mass.
Therefore, we hypothesize that the difference between our and their results is mainly due to the significant differences between the methods used to fit the data: on one hand,
a non parametric FSF method based on a linear combination of SSPs to fit the spectrum; 
on the other hand, a parametric FSF based on constrained SFHs to fit the spectrum and the photometric data. 

At redshift comparable to VANDELS data, we show in Fig. \ref{fig:mass_Z} the mass-weighted 
metallicity values derived by \cite{kriek19} for 5 massive (log(M$_*$/M$_\odot$)$>$10.6) 
galaxies at $z\sim1.4$, the metallicity derived by \cite{onodera15} from the stacked 
spectra of 16 passive galaxies at $z$$\sim$1.6 and the metallicity estimate
by \cite{lonoce15} for a massive galaxy at $z$$\sim$1.4.
The measurements by \cite{kriek19}, based on the absorption lines fitting code and models
developed by \cite{conroy12}
\citep[see also][]{conroy14,choi14}, agree with the positive correlation of the metallicity 
with mass.
The measurement by \cite{lonoce15} is based on absorption lines fitting with M11 models \citep{maraston11} and, taken together with the \cite{kriek19} estimates, 
confirm the large scatter we also observe in the stellar metallicity of individual galaxies \citep[see also][]{lonoce20}.
{ It is worth to mention that large differences are found by \citet{spiniello12}
between predictions of some line indices of \cite{conroy12} models with respect to \cite{vazdekis15} models.}
The metallicity derived by \cite{onodera15}, based on the comparison of absorption
line indices with the models' predictions by \cite{thomas11}, is consistent with the 
metallicity values we derived at comparable mass.

At higher redshift, we show the mass-weighted metallicity derived by \citet{kriek16}
for a massive quiescent galaxy at $z$=2.1\footnote{As far as we know,
there are no other metallicity estimates for quiescent galaxies at $z$$>$2.5
besides the estimate at $z$$\simeq$3.35 by \citet{saracco20apj}.
Recently, \cite{cullen19} and \cite{calabro21} derived stellar metallicities for 
VANDELS starforming  galaxies at $z$$>$2.5 by comparing the far UV 
($\lambda_{rest}$$<$2000 \AA) spectral features of stacked spectra with the theoretical 
stellar library of massive stars by \cite{leitherer10}.}
according to the same method used in \citet{kriek19}.
The metallicity agrees with those derived for galaxies of similar mass both at lower
and higher redshift.

Therefore, the different estimates from the literature, with the exception of those
by \citet{carnall19,carnall22},  agree with our estimates: 
the average stellar metallicity of massive (log(M$_*$/M$_\odot$)$>$11) passive galaxies is supersolar, 
higher than for lower mass galaxies, and it did not change across cosmic time, at least
over the last 9 Gyr.
This lack of a significant evolution in the mean metallicity value seems to apply to 
the whole population of passive galaxies in the mass range 
10$<$log(M$_*$/M$_\odot$)$<$11.6, as suggested by the agreement among the different 
estimates in the redshift range 0$<$$z$$<$1.4.
The detection at $z$$\sim$1.2 of a positive correlation between metallicity and mass,
consistent with the one observed in the local Universe, shows that the observed
trend was established at earlier epochs as result of the formation process rather than their evolution.

\section{The star formation history} 
\subsection{The formation epoch of stellar mass}
\begin{figure}	
	\includegraphics[width=8.5truecm]{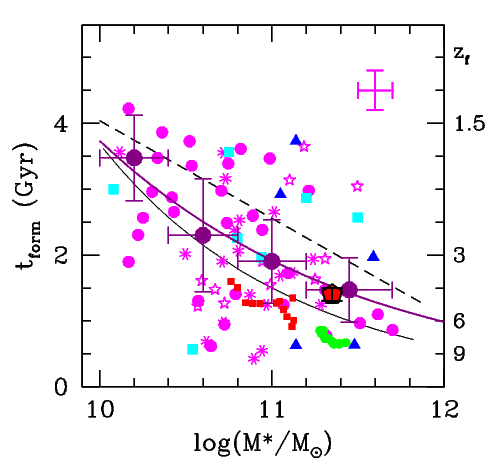}
   \caption{\label{fig:tf_mass} Formation epoch/redshift (left/right axis) as a function 
   of stellar mass.
   Symbols are as in Fig. \ref{fig:mass_Zevol}.
   The purple filled circles are the median t$_{form}$ values of VANDELS galaxies in the different
   mass intervals as marked by horizontal error bars.
   Vertical error bars represent the standard deviation of the values in each mass bin.
  The cross in the upper right corner represents the typical error at 1$\sigma$ of individual
  values.
  The purple thick solid curve is the best fitting $t_{form}$-M$_*$ relation derived for
  the median t$_{form}$ values of VANDELS galaxies (see text, Eq. \ref{eq:tf}).
  The black dashed line is the best fitting $t_{form}$-M$_*$ relation found
  by \citet{carnall19}.
  The black thin solid curve is the $t_{form}$-M$_{dyn}$ relation derived by \citet[][Eq. 8]{saracco20} from the FP 
  { scaled according to the mean value $\langle$M$_{dyn}$/M$_*$$\rangle$$\simeq$1.6 they find for their sample}.
  }
\end{figure}

Figures \ref{fig:comparison} and \ref{fig:comparisonUV} (lower panels)
show the stellar age resulting from the fitting to stacked spectra as a function
of their mass.
A mild increase of the age with the mass can be seen, regardless of the choice of the models.
However, galaxies that contribute to the stacked spectra in each mass interval are spread 
over the redshift range 1.0$<$$z$$<$1.4, corresponding to an interval of time 
$\Delta$t$\sim$1.3 Gyr which could affect the real trend.
To properly compare the mean stellar age of galaxies seen at different redshift, we 
considered the mean formation epoch of the stellar mass defined as 
$t_{form}$=Age$_U$($z$)-Age$_{M*}$($z$), where Age$_U$($z$) is the age of the Universe 
at the redshift of the galaxy and Age$_{M*}$($z$) is its mass-weighted age. 

Fig. \ref{fig:tf_mass} shows the mean formation epoch $t_{form}$ of the stellar mass as a 
function of mass for the VANDELS sample and the other passive galaxies at similar and higher 
redshift as well as those derived for stacked massive ETGs at lower redshift.
Large purple filled circles are the median $t_{form}$ values of VANDELS passive galaxies
in the different mass intervals.
The well known general trend between formation epoch and stellar mass is clearly visible from the figure: the higher the stellar mass of a galaxy the earlier formed (the older are) 
its stars, in agreement with previous studies of stellar population properties in local 
galaxies \citep[e.g.][]{cowie96,kauffmann03,gallazzi05, thomas05, thomas10,conroy14,mcdermid15} and of
scaling relations for local and higher redshift galaxies 
\cite[e.g.,][]{mcdermid15,barone18,saracco20}.
For comparison, Fig. \ref{fig:tf_mass} shows the $t_{form}$-M$_*$ relation found by \citet[][dashed line]{carnall19} on VANDELS passive galaxies, and the relation derived
by \citet[][thin solid curve]{saracco20} from the study of the Fundamental Plane of 
cluster ETGs at $z$$\sim$1.2.
The relations follow the same trend with similar slopes even if they are
 offset by about 1.0 Gyr.
By fitting the median $t_{form}$ values of VANDELS galaxies as a function of stellar
mass we derived the relation:
\begin{equation}
\label{eq:tf}
 log(t_f/Gyr)=(-0.29\pm0.04) \times log(M_*/M_\odot)+(3.5\pm0.4)
\end{equation}
represented by the thick purple solid curve which lies in between the two 
previous relations.

\subsection{Star formation and stellar mass}
The $t_{form}$-M$_*$ relation expressed by Eq. \ref{eq:tf} summarizes the relationship 
between SFH and mass.
From the fit to VANDELS stacked spectra (Fig. \ref{fig:stack}), the SFH seems to be 
smoother and longer (i.e., star formation episodes 
distributed over a larger interval of time) in lower mass galaxies, while is sharper 
and shorter (i.e., characterized by one or two major episodes accounting for more than 50 
per cent of the mass) 
in higher mass ones.
In these latter, more than 10$^{11}$M$_\odot$ formed within $\sim$1 Gyr.

\begin{figure*}	
	\includegraphics[width=8.5truecm]{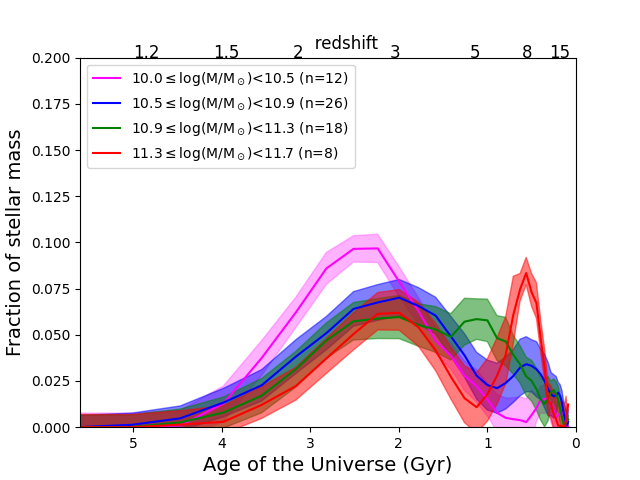}
 	\includegraphics[width=8.5truecm]{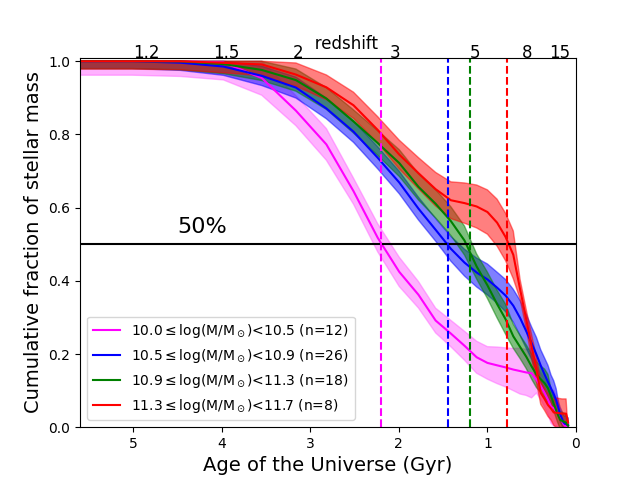}
   \caption{\label{fig:sfh_cumul} Left - Average SFH of VANDELS passive galaxies in different
   ranges of stellar mass as a function of cosmic time. 
   Continuous curves are the mean mass fraction as resulting from the best fitting composite model of individual galaxies for different range of stellar masses.
   The dashed regions indicate the dispersion of the values divided by $\sqrt{N_{gal}}$
   at each time interval. 
   Right - Average integrated SFH showing the cumulative mass fraction existing at 
   a given time. 
   The vertical dashed lines marks $t_{50}$ ($z_{50}$), the age of the Universe (redshift, 
   top x-axis) at which 50 per cent of the stellar mass of galaxies for the different mass ranges was already formed. }
\end{figure*}
Fig. \ref{fig:sfh_cumul} (left) shows the mean SFH of VANDELS galaxies in different mass bins, i.e., the mean fraction of stellar mass as a function
of time as resulting from the fitting to individual galaxies,
instead of considering stacked spectra as in Fig.\ref{fig:stack}. 
The right panel shows the mean integrated SFH, the cumulative fraction of mass.
The average SFH has been obtained according to the following procedure.
For each galaxy, we considered the fraction of stellar mass associated to each SSP 
contributing to the composite best fitting model and, for each of them, we derived
the corresponding $t_{form}$ from the age of the SSP.
We then re-sampled and summed the fractions for all the galaxies within 
intervals of log(d$t$[Gyr])=0.05 \citep[see also][]{asari07}.
Finally, we run a moving mean (MM) over 5 log(d$t$) to smooth the data.
We verified that the results do not depend on the binning assumptions by
repeating the procedure for different binning.
We note that quite different SFHs exist
among galaxies with similar mass \citep[see also, e.g.,][]{carnall19,tacchella22}.

Fig. \ref{fig:sfh_cumul} shows that at increasing stellar mass the fraction 
of stars that formed at early epochs increases: the SFH tends to be more skewed
toward early epochs.
Less massive galaxies host younger stars whose formation started later, according to a SFH
peaked at more recent epochs than for more massive galaxies.
Moreover, Fig. \ref{fig:sfh_cumul} also suggests the presence of a double peaked SFH.
The relative intensity of the peaks seems to correlate with the mass: 
the higher the mass the more pronounced the first peak at early epochs. 
{ Whether the double peak is real or simply the result of the discrete nature of the SFH 
derived here (sum of the SSPs), the formation of stellar mass begins earlier and earlier 
as the mass of the galaxy increases more and more}.
Indeed, the right panel shows that, on average, the stellar mass in 
higher mass galaxies is formed earlier than in lower mass ones,  as shown by the epoch 
at which 50 per cent of all stars in a galaxy is already formed \citep[see also, e.g.,][]{estrada20}.
More than 50 per cent of the stellar mass in massive 
log(M$_*$/M$_\odot$)$>$11.3 passive galaxies is formed
at $z$$>$5, and almost 80 per cent within the first 2 Gyr of the cosmic time, i.e. 
by $z$$\sim$3.
This agrees with the results shown in Fig. \ref{fig:stack} 
derived from the fit to the stacked spectra.

Significant star formation took place at early epochs.
Indeed, galaxies with masses log(M$_*$/M$_\odot$)$>$10.5 experience an epoch of 
significant star formation at the earliest cosmic times, 
as shown by the growth of stellar mass within the first 1.5 Gyr.
However, at $z$$\sim$1.2, these galaxies show quite different properties,
in terms of age and metallicity.
These differences are mainly driven by the different SFHs and by the
likely different rate at which star formation decline. 
We take up this aspect in more detail in the next section.

These results are qualitatively similar to those derived for local early-type galaxies
\citep[e.g.,][]{thomas05,thomas10,mcdermid15}.
{We underline the fact that the results obtained here are independent of any assumption.}
The power of the non-parametric full spectral fitting, as the one adopted in this analysis, 
is just to detect differences in the SFH resulting from differences in the spectral properties.
All the galaxies can indeed be modeled by linearly combining SSPs extracted from the same 
base of models without any prior, apart from the discretization of the models (the same
for all the galaxies).
{ Therefore, the fact that stellar populations in galaxies of greater mass are, on average, described by different SFHs than stellar populations in galaxies of lower mass reflects precisely the presence of systematic differences in their spectral and, therefore, stellar properties.} 

\section{Stellar Metallicity and SFH}
\begin{figure*}	
	\includegraphics[width=8.5truecm]{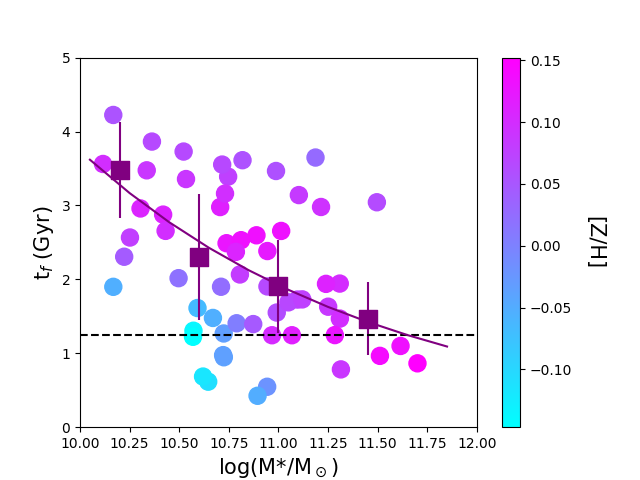}\hskip 1truecm
	\includegraphics[width=8.5truecm]{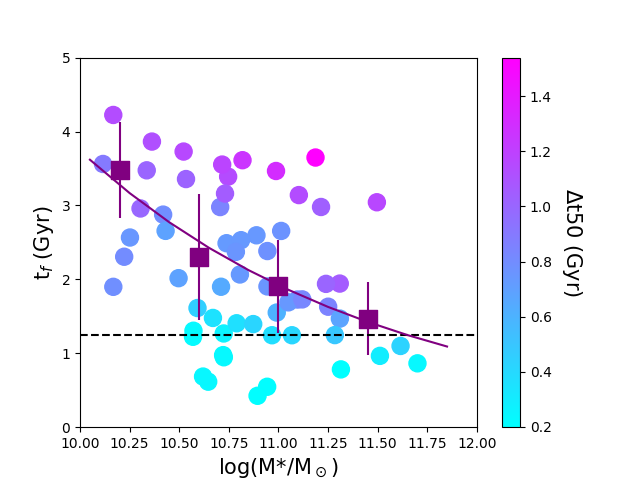}
	\includegraphics[width=8.5truecm]{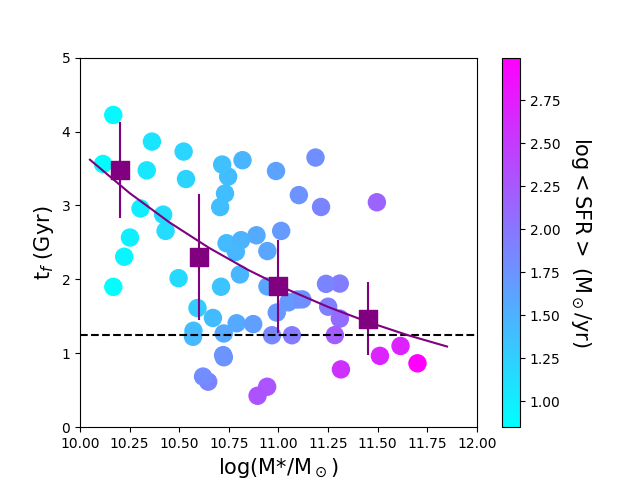}	
   \caption{\label{fig:4plots_zh} Formation epoch t$_{form}$ as a function of stellar 
   mass for the VANDELS galaxies (filled circles). Colorscale represent LOESS-smoothed mass-weighted metallicity [Z/H] (upper panel), time interval $\Delta t_{50}$ within which 
   50 per cent of the stellar mass is formed (lower left; see text) and SFR averaged over 
   the time required to form 90 per cent of the stellar mass (lower right).  The median 
   t$_{form}$ values of VANDELS galaxies in the different mass intervals (purple squares) 
   and the best fitting $t_{form}$-mass relation (purple curve; Eq. \ref{eq:tf}) 
   are also shown. The dashed line defines Maximally Old Galaxies (MOGs, see \S\ 6), those
   hosting stellar populations formed at $z$$>$5, within the first $\sim$1.2 Gyr of 
   the cosmic time.}
\end{figure*}

\begin{figure*}	
    \includegraphics[width=8.5truecm]{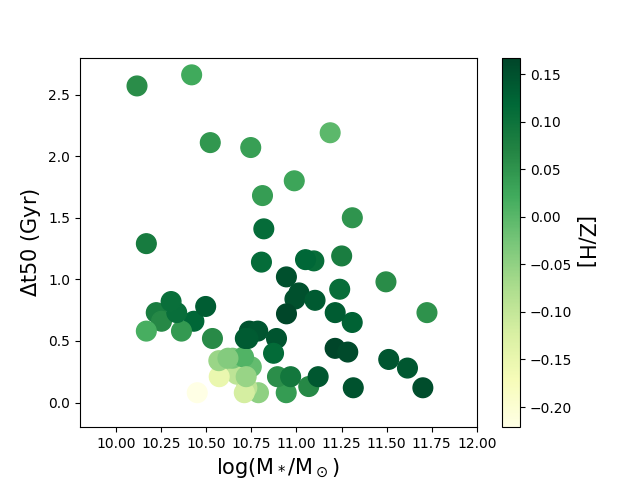}
	\includegraphics[width=8.truecm]{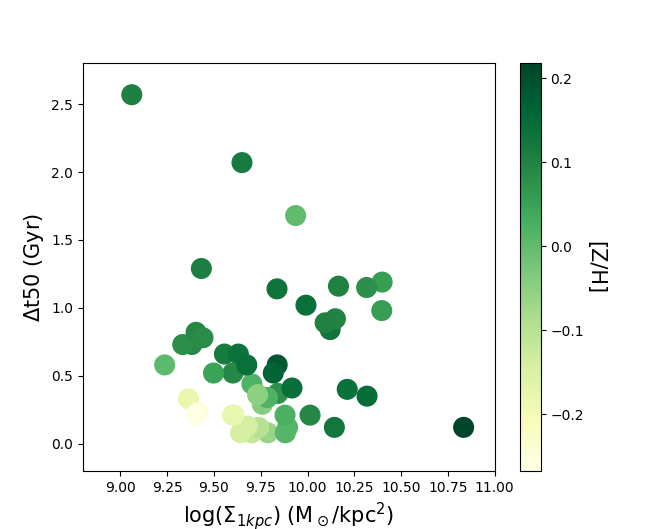}
   \caption{\label{fig:dt50_rho1} Time $\Delta$t50 required to form 50 per cent of the mass 
   as a function of mass (left) and central stellar mass density $\Sigma_{1kpc}$ (right).  Colorscale represents LOESS-smoothed mass-weighted metallicity [Z/H]. { Note that right hand panel shows 51 passive galaxies out of the 64: 25 galaxies for which an effective radius 
   (R$_e$) was available from the literature \citep[][]{vanderwel12, vanderwel14} 
   and 26 for which R$_e$ has been derived from HST-F160W archive images.}}
\end{figure*}
Fig. \ref{fig:4plots_zh} shows how the stellar metallicity, the time scale to form half
of the stellar mass  
and the mean SFR (see below) are distributed among galaxies on the $t_{form}$-M$_*$ plane. 
The upper panel shows the distribution of the mass-weighted metallicity 
[Z/H]. 
The colorscale shows [Z/H] smoothed using the locally weighted 
regression algorithm LOESS \citep{cleveland88, cappellari13} to
highlight the presence of possible trends.
The figure joins the relationships between metallicity and age with the mass
of galaxies, 
correlations individually seen in figures \ref{fig:mass_Z} and \ref{fig:tf_mass},
respectively.
Even if the statistic is low, this 3D view adds an information that was not
evident from the two individual figures: contrary to high-mass 
galaxies (log(M$_*$/M$_\odot$)$>$11), that all have supersolar metallicity and
whose stellar mass formed earlier (old stellar ages), 
the stellar metallicity of lower mass galaxies (e.g., log(M$_*$/M$_\odot$)$\sim$10.6) 
is sub-solar if formed at earlier epochs (old ages), supersolar if formed at later epochs
(young ages).
{ Invoking the age-metallicity degeneracy to justify this trend is of no help:
high mass galaxies, all formed at earliest epochs, i.e. hosting
oldest stars, should be all metal poor, contrary to what it is found. } 

Focusing on galaxies hosting the oldest stellar population, that is those with 
$t_{form}$$<$1.2 Gyr ($z_{form}$$>$5) that we call maximally old (MOGs)\footnote{ The definition of MOGs is arbitrary. 
The selection $t_{form}$$<$1.2 Gyr follows the choice of considering the $\sim$1 Gyr of
SF between the first star forming events possibly taken place at $z$$\sim$15-20 
(cosmic time $\sim$0.2 Gyr), and the end of the re-ionization ($z$$\sim$5-6).} hereafter,
the positive trend between [Z/H] and mass is extremely clean.
By selection, the stellar populations in these galaxies formed nearly within the same 
(short) time. 
However, their metallicity is significantly different and increases systematically 
with the mass of the galaxy.
Conversely, focusing on low mass galaxies, at fixed mass say, e.g.
log(M$_*$/M$_\odot$)$\sim$10.6, the metallicity is not constant for all of them
but increases systematically with $t_{form}$ and reaches supersolar values as high 
as for the most massive galaxies.
Therefore, the trend between metallicity and mass cannot be the only causal relationship 
but there must be other processes that overlap and modulate this relationship.

In the lower-left panel the colorscale shows the quantity $\Delta{t50}$ (Gyr) defined as
the time to form 50 per cent of the stellar mass of the galaxy,
starting when the star formation begins.  
It is the time needed to form half of the mass (a proxy for the duration of the SF) 
derived from the SFH of galaxies  described in the previous section.
We remark that $\Delta{t50}$ differs from $t_{50}$ shown in Fig. \ref{fig:sfh_cumul} 
since the latter marks the cosmic epoch at which 50 per cent of the stellar mass formed 
while the former indicates the time required to do it.
Correctly, MOGs are characterized by similar $\Delta{t50}$.
On the contrary, the systematic increase of the metallicity with $t_{form}$ at fixed 
(low) mass, is accompanied by increasing values of $\Delta{t50}$: in low mass galaxies,
shorter SF produces lower stellar metallicity, longer SF higher metallicity.
We notice that while low mass galaxies can be described by both long and short 
values of $\Delta_{t50}$, high-mass galaxies are described only by short 
values.
{ In fact, we find that $\Delta{t50}$ is anti-correlated with the mass (and mass density)
as shown in Fig. \ref{fig:dt50_rho1}, i.e. the time scale of star formation depends on the mass 
(and mass density) of galaxy, as found for local ETGs (e.g., \citet{thomas10,mcdermid15}; but see \citet{beverage21} for a different result)}.

In the lower-right panel of Fig. \ref{fig:4plots_zh} the colorscale shows the average  
$\langle SFR\rangle$ (M$_\odot$/yr)
defined as the ratio between 90 per cent of the mass and the time interval to form it.
As expected, $\langle SFR\rangle$ increases with the mass, { showing that the rate at 
which gas is converted into stars increases faster with mass than the duration
of the SF.}
At fixed mass, $\langle SFR\rangle$ increases with decreasing $t_{form}$ since 
decreases also $\Delta t_{50}$.
Focusing on MOGs, given that they all have approximately the same $\Delta t_{50}$, $\langle SFR\rangle$
scales simply according to the mass, from $\sim$30 M$_\odot$/yr at 
log(M$_*$/M$_\odot$)$\sim$10.6 to more than 300 M$_\odot$/yr at log(M$_*$/M$_\odot$)$\sim$11.6.

Fig. \ref{fig:dt50_rho1} shows $\Delta{t50}$ 
as a function of mass and central stellar mass density $\Sigma_{1kpc}$=${M}_{1kpc}$/($\pi R_{1kpc}^2)$, defined as the mass density within 1 kpc radius \citep{saracco12,saracco17}\footnote{ The stellar 
mass interior to 1 kpc is given by
\begin{equation}
\mathcal{M}_{1kpc}={L_{1kpc}}\times\left ({\mathcal{M}_*\over L_{tot}}\right )_{gal}
={\gamma(2n,x)\over \Gamma(2n)}\times\mathcal{M}_*
\end{equation}
where 
\begin{equation}
L_{1kpc}=2\pi I_eR_e^2  n {e^{b_n}\over{b_n^{2n}}}\gamma(2n,x)
\label{l1}
\end{equation}
is the luminosity within the central region of 1 kpc, 
L$_{tot}$ is the total luminosity obtained by replacing in Eq. \ref{l1} 
$\gamma(2n,x)$ with the complete gamma function $\Gamma(2n)$
\citep{ciotti91}, $\mathcal{M}_*$ is the stellar mass of the 
galaxy, $n$ is the S\'ersic index, $x=b_n(R_1/R_e)^{1/n}$ and 
$\gamma(2n,x)$ is the incomplete gamma function.
We assumed the analytic expression $b_n=1.9992n-0.3271$ 
\citep{capaccioli89} to approximate the value  of $b_n$. }.
The colorscale shows the metallicity [Z/H].
The figure shows a trend between $\Delta{t50}$ and $\Sigma_{1kpc}$ recalling the one 
in the upper panel of Fig. \ref{fig:4plots_zh}:
highest densities are associated with shortest duration and highest metallicity; 
lower densities are associated both with short and long duration with the metallicity 
that increases as $\Delta{t50}$ increases.

Taken together, these relations suggest that the duration of the SF (the time scale 
to form half the mass, $\Delta{t50}$), and the rate at which the gas is converted 
into stars ($\langle SFR\rangle$) play a role in the definition of the 
stellar metallicity depending on the mass and/or on the mass density of the galaxy.
We discuss further these results in the Discussion.

In Appendix \ref{sec:indices}, we derive spectral indices (and make them publicly available)
and use them to qualitatively test, in a model independent way, the relationship between SFH, age, metallicity and mass, as derived
with npFSF.
We also use Mg and Fe UV lines in an attempt to constrain [$\alpha$/Fe] abundance ratio.
{  We remind that the relative abundances of $\alpha$ elements and iron
would provide crucial information in terms of star formation timescales \citep[e.g.][]{tinsley79, matteucci94}, and insights on possible variations of the IMF \citep[e.g.,][]{labarbera13}. 
Unfortunately, our analysis cannot be conclusive due to the still poor characterization of features in the UV spectral range. 
High S/N near-IR observations enabling to extend the spectral coverage
up to $\lambda_{rest}$$\sim$5800 \AA\ would be needed (see e.g., Bevacqua et al.,
in preparation).}

\section{Summary of the results}
In this work we studied the stellar population properties of early-type and passive
galaxies in the mass range 10$<$log(M$_*$/M$_\odot$)$<$11.6 at 1.0$<$$z$$<$1.4 using 
VANDELS data.
We first investigated the dependence of the age and metallicity estimates 
on the methods and assumptions usually adopted in this kind of analysis.
This part of analysis showed that:
\begin{itemize}
 \item
Significant systematics exist among estimates from different stellar population 
models, different spectral ranges as often happens for samples at different redshifts, 
and when luminosity-weighted or mass-weighted quantities are considered.
These systematics highlight the need for a uniform and homogeneous methodology at all redshifts.
\end{itemize}
On the basis of these results, the stellar population properties of galaxies have been 
derived using homogeneous measurements and method over the whole redshift range 
considered, i.e., same fitting code (STARLIGHT), simple stellar population models
(EMILES) and fitting wavelength range ([3350-4350]\AA).
We thus defined the age-mass and the metallicity-mass relations at these redshifts 
and studied the evolution of the stellar metallicity down to $z$$\sim$0.
We then reconstructed the SFH to study its relation with stellar population 
and physical properties of ETGs, and to trace the origin of the stellar MZR and its evolution.
The main results are the following:
\begin{itemize}
 \item The stellar metallicity and age of VANDELS passive galaxies at $\langle z\rangle$$\sim$1.2 are 
 positively correlated with their stellar mass, as for galaxies in the local Universe:
 higher mass galaxies host stars formed earlier and are more metal rich than most of the
 lower mass galaxies (see \S 4.1).
 These trends are, in a relative sense, independent of models or any particular assumption, as also confirmed 
 by the observed trends of age and metallicity sensitive absorption spectral indices
 with mass (see Appendix).
 { However, different models might change the absolute values one infers, as shown
 in Sec. 3.} 
It is worth to remind the reader that these correlations are mainly due to the increasing 
lack of young and low-metallicity galaxies as the mass increases, an effect clearly 
present also in local \cite[e.g.,][]{gallazzi05,mcdermid15,barone18} and intermediate redshift 
 \cite[e.g.,][Bevacqua et al., in preparation]{barone22} samples.
 \item The stellar metallicity of VANDELS passive galaxies with mass 
 10$<$log(M$_*$/M$_\odot$)$<$11.6 falls in the range -0.35$<$[Z/H]$<$0.26\footnote{ We stress
 that [Z/H]=0.26 is the upper limit of the metallicity of the EMILES models considered
 here. Therefore, we cannot rule out metallicity values even higher than this one for some massive galaxies.},
 with 67 per cent of them having [Z/H]$>$0.0.
 This percentage rises to 90 per cent (19/21) at masses log(M$_*$/M$_\odot$)$>$11.
 These metallicity values agree with those estimated for field ETGs
 with comparable mass at $z$$\sim$1.4 from \cite{gargiulo16}, with 
 those in the cluster XLSS0223 at $z$$\sim$1.2 from \cite{saracco19} and with the few
 estimates at similar redshift from the literature \citep{kriek19, onodera15, estrada19}.
 On the contrary, for galaxies with mass log(M$_*$/M$_\odot$)$\sim$11.2 we estimate a mean stellar metallicity ([Z/H]=0.17$\pm$0.1) $\sim$2$\sigma$ higher than the metallicity 
  estimated by \cite{carnall19} for VANDELS passive galaxies with 
 the same redshift and mass 
 { ([Z/H]=-0.07$\pm$0.08 when corrected by the offset of +0.06 between models, 
 see note \ref{foo:offset})}. 
 We verified that this significant difference is not due to any of the possibile systematics
 introduced by different SSP models, spectral range or quantities considered. 
 We believe that the difference is due to the very different methods adopted 
 by the used fitting codes (see \S 4.3).
  \item We do not detect any cosmic evolution of the metallicity-mass relation, 
  either in the slope or in the normalization down to $z$$\sim$0, as confirmed
 by the comparison with the relation derived from LEGA-C passive galaxies
 at $z$$\sim$0.65 and with the local relations from the literature \citep{gallazzi14,choi14,peng15}.
 { This confirms that the stellar metallicity is mainly defined during the formation
 of the dominant (in mass) stellar population. 
 According to EMILES models, massive galaxies (log(M$_*$/M$_\odot$)$>$11) have 
 supersolar metallicity  
 and subsequent evolutionary processes (merging and/or SF, those processes able to modify 
 the macroscopic properties of a galaxy)} do not modify it (see \S 4.2).
  \item
  The comparison of Mg and Fe UV spectral indices of VANDELS stacked spectra 
  with those of stacked massive ETGs at $z$$\sim$0.38 does not {allow us
  to reach a firm conclusion about the possible evolution of
  the [$\alpha$/Fe] ratio with redshift} (see Appendix).
 \item
 The cumulative SFHs of VANDELS passive galaxies show that the fraction of stellar mass formed at 
 early epochs increases with the mass of the galaxy, in agreement with the positive 
 age-mass relation.
 On average, about 80 per cent of the stellar mass of very massive (log(M$_*$/M$_\odot$)$>$11.3) 
 galaxies formed within the first 2 Gyr of cosmic time ($z$$>$3), and 50 per cent within
 the first Gyr (by $z$$\sim$5), results qualitatively in agreement with those derived 
 for local early-type galaxies
 \citep[e.g.,][; see \S 5]{thomas05, thomas10, mcdermid15}.
 \item 
 Massive galaxies (log(M$_*$/M$_\odot$)$>$11.0) host old stellar populations
(t$_{form}$$<$2 Gyr) characterized by supersolar metallicity ([Z/H$>$0.05).
These stars have been formed in short time ($\Delta$t50$<$1 Gyr) implying 
high star formation rates (SFR$>$100 M$_\odot$/yr) originating in high mass density 
regions, log($\Sigma_{1kpc}$)$>$10 M$_\odot$/kpc$^2$.
 This sharp picture tends to blur with decreasing mass:
 galaxies with intermediate mass, e.g., log(M$_*$/M$_\odot$)$\sim$10.6 can host 
 either stars with 
 sub-solar metallicity as old as those in massive galaxies, or younger stars with 
  supersolar metallicity, depending on the duration of the 
  star formation, shorter or longer respectively (see \S 6;), in agreement
  with other studies at intermediate redshift (Bevacqua et al., in preparation).
\end{itemize}

\section{Discussion and conclusions}
To study the evolution of the stellar populations properties of galaxies, 
in particular of metallicity, it is essential to compare estimates obtained 
in a homogeneous way at the different redshifts.
Our analysis showed, indeed, that such kind of estimates must be homogeneous to be comparable and that a metallicity variation (if any) should be considered relative since values,
being dependent on the considered models, methodology and spectral range,
are not absolute.

The fact that positive correlations between stellar age and metallicity with the mass of 
the galaxies were already in place at redshift $z$$>$1.2 implies that they were defined 
during the first 4 Gyr of cosmic time, as a result of the formation processes
of stellar mass and galaxies. 
The lack of evolution in stellar metallicity of galaxies with log(M$_*$/M$_\odot$)$>$10.6 
in the last 9-10 Gyr (0$<$$z$$<$1.4) and, perhaps, $\sim$12 Gyr ($z$$<$3.35) confirms on 
the one hand that evolutionary processes (merging and/or later SF) do not significantly influence the stellar 
metallicity of galaxies, at least of the massive ones, and that the metallicity of 
galaxies is determined at early epochs, during the main massive star formation event.

First, let's consider the constraints on the evolution imposed by the lack of 
metallicity evolution over the last $\sim$9 Gyr. 
Once a given stellar mass is seen assembled in a galaxy (progenitor) with a given age and metallicity [Z/H], to generate a descendant with a metallicity, e.g., 
0.1dex lower (higher), a similar mass with metallicity, at least, half (twice) of the progenitor must be added.
The lower the accreted mass fraction, the higher its metallicity offset with respect
to the progenitor and vice versa.
The same reasoning applies to the mean metallicity of the population of galaxies: to change
it by $\Delta$[Z/H]=$\pm$0.1 a similar amount of galaxies (with similar mass distribution)
with metallicity half or twice should be added to the population.
Therefore, a lack of metallicity evolution especially for 
masses $\sim$10$^{11}$ M$_\odot$ or higher is not surprising.
New star formation of this magnitude is not seen at $z$$<$1.4 \citep[e.g.,][and references therein]{madau14}, the mass growth of massive galaxies in this redshift range, if any,
is negligible \citep[e.g.,][]{muzzin13} and, most importantly, the stellar populations in 
local massive ETGs are, on average, old as expected for passive evolution
\citep[e.g.][]{gallazzi05, thomas05}.
Major merging (mass ratio $\sim$1:1) would leave the metallicity unchanged since it 
is expected that similar mass galaxies have similar stellar metallicity.
Minor mergers would not affect the mean stellar population properties
of a massive progenitor given the low mass (expected metal-poor) accreted, even if 
they could efficiently affect the structural properties of a galaxy 
\citep[e.g.][]{ciotti07, hopkins09, naab09, bezanson09}\footnote{Major and minor mergers can play a role in the evolution of galaxies even if in a limited way because of the large scatter that they would introduce
in the scaling relations \citep[see e.g.,][]{nipoti09}.
We do not discuss further this topic since it is out of the scope of this analysis.}.
For instance, \cite{hirschmann15} showed that minor mergers can steeping the stellar
metallicity gradients due to the accretion of metal-poor stars at the outskirts of massive
galaxies, leaving unchanged their mean metallicity.

Newly formed massive galaxies can play a role in the evolution of the mean
properties of the population of early-type and passive galaxies at different redshifts \citep[progenitor bias, e.g.][]{fagioli16}.
However, considering the arguments above, their stellar population properties  cannot
be substantially different from galaxies already assembled, apart from their size and
mass density.
For these reasons, and independently of our result, it is unclear how the fall 
by $\sim$0.3dex in the mean stellar metallicity of galaxies with log(M$_*$/M$_\odot$)$>$11.2 
found by \cite{carnall22} at $z$$\sim$1.2 with respect to the local Universe can be justified,
unless we assume that local estimates are all biased (for some reason) 
toward metallicity values $\sim$2 times higher than the value they find at that redshift.
{ A comparison among estimates homogeneously obtained with their methodology  
on different samples down to the local Universe (as we did with the method adopted 
in this work) would clarify whether the strong evolution of the metallicity they found is 
due to the different methodology used in the comparison estimates.}

Much more indicative than the lack of metallicity evolution is the nature of the 
stellar MZR. 
Our analysis shows an increasing lack of young
and low-metallicity galaxies as the stellar mass increases: there are no massive galaxies 
as young as some of the low-mass galaxies, and no massive galaxies with metallicity 
as low as some of the low-mass galaxies.
On the contrary, low-mass galaxies can display either supersolar metallicity, as high as massive ones, associated with young ages, or sub-solar metallicity associated with old ages. 
This is evident in our analysis, for galaxies at $z$$\sim$1.4, in those at intermediate
redshift (see e.g. Bevacqua et al., in preparation) and in the local Universe
\citep[see e.g.,][]{gallazzi05,asari09,peng15}.

Given the redshift of our galaxies, these differences must be the results of
different SFHs and/or initial conditions.
The fact that the most massive galaxies host old stellar populations characterized by 
supersolar metallicity, formed within short time and high SFR associated with high
central stellar mass densities are all properties expected for stellar systems that 
form in the early Universe \citep[e.g.,][]{wellons15}.
The higher mean density of the early Universe favors higher molecular gas densities and, according to the Kennicutt-Schmidt law \citep{kennicutt98}, higher star formation rate densities.
The high SFR determined by the short time and the high mass, efficiently enriches the 
interstellar medium \citep[e.g.][]{matteucci94, calura09}.
This enhances the stellar metallicity more efficiently than in lower mass galaxies
even in case of outflows
\citep[at first guess, proportional to the SFR; e.g.,][]{calura09, spitoni10, spitoni17}. 
Indeed, low-mass galaxies are expected to have a lower ability than high-mass ones 
in retaining the metals because of their shallower potential well 
\citep[e.g.][]{dekel86, tremonti04, delucia04, kobayashi07, finlator08}.
This is confirmed by the observed correlation between velocity dispersion, 
a direct measure of the potential well, and stellar metallicity 
\citep[e.g.][]{jorgensen99,trager00,harrison11,mcdermid15}.

For short SF timescales, such as those of massive galaxies, low-mass galaxies show
much lower stellar metallicities associated with old ages.
The SFR is low given the low mass and the low density, the enrichment of the ISM is 
less efficient than in high-mass galaxies and requires much longer time to enhance 
the stellar metallicity.
Indeed, for low-mass galaxies, metallicity increases as the duration of the SF increases
and it is, therefore, associated to younger stellar population.
\cite{calura09} show that a variable star formation efficiency from low- to high-mass
galaxies (higher in massive galaxies) can produce the observed stellar MZR.
\cite{spitoni10} conclude that a plausible scenario is variable star formation efficiency 
coupled to galactic winds becoming more important in low-mass galaxies
\cite[see also][]{spitoni17}.
We cannot exclude that SF efficiency changes systematically with mass, even if it is not
clear what physical mechanism can change the SF efficiency. 
{ In this respect, it is worth to remind that GAEA correctly reproduces the shape
of the stellar MZR for passive galaxies at $z$$<$1.4 as well as the lack of evolution
at large stellar masses without assuming any scaling of the SF
efficiency with stellar mass.}

To conclude, our results show that some properties characterizing the SFH
change systematically among galaxies and that these variations underlie the MZR.
In particular, we find evidence that the SFR and the duration of the SF are the 
properties playing a major role in defining the stellar MZR and age-mass relations of 
ETGs with the fundamental complicity of the underlying mass which modulates the 
retention of metals as the mass increases.

\section*{Acknowledgements}
Based on public data products from observations made with ESO Telescopes at the La Silla or 
Paranal Observatories under programs ID 194.A-2003(E-K) and ID 194.A-2005(A-F).
This work is partially based also on observations carried out with the ESO Very Large
Telescope under programme ID 085.A-0135, with the Large Binocular Telescope (LBT)
under programs ID 2015-2016-28 and ID 2017B-C2743-3.
PS, FLB, RDP, DB, FF, AP, CS and CT acknowledge support by the grant 
PRIN-INAF-2019 1.05.01.85 ($\#$ 11). 
CS is supported by an `Hintze Fellow' at the Oxford Centre for Astrophysical Surveys, 
which is funded through generous support from the Hintze Family Charitable  Foundation.
PS would like to thank V. Lynd for the useful discussions.
PS thanks Anna Gallazzi and Stefano Zibetti for their careful reading and
comments on the manuscript.


\section*{Data availability}
The VANDELS data underlying this article are available at http://vandels.inaf.it/ or
via the ESO Archive (https://www.eso.org/qi/).
The codes \texttt{STARLIGHT} and \texttt{pPXF} are publicly available.
The spectral indices generated in this research are published in electronic form 
(Tables B1 and B2).
The remaining derived data will be shared on reasonable request to the corresponding author.

\bibliographystyle{mnras}
\bibliography{pap_vandels1_rev1} 




\appendix
\section{Simulating mixed stellar populations}
 \label{sec:appsimul}
{  We checked whether and how the spectral 
range [2600-3350] \AA\ affects the results of the full spectral fitting 
performed over a range extending up to $\lambda_{rest}$$\leq$4350 \AA,
in case of multiple stellar populations.
To this end, we simulated a galaxy (a template) with a mixed stellar population starting from EMILES SSP models.
The simulated galaxy has the bulk (weight=1) of the stellar mass represented by
a SSP 3 Gyr old.
To this stellar population we added 10\% of stellar mass (weight=0.1)
represented by a SSP 0.2 Gyr old.
Both the SSPs have metallicity [Z/H]=0.15.
Therefore, the resulting stellar population is 2.75 Gyr old and has a 
metallicity [Z/H]=0.15.

This template has been normalized to the VANDELS stacked spectrum
with log(M$^*$/M$_\odot$)=11 in the range 4100-4250 \AA.
Then, the residuals of the full spectral fitting to the stacked spectrum have
been randomly reshuffled 30 times in $\lambda$ and summed to the 
template to reproduce similar S/N and mimic the true uncertainties.
In Fig. \ref{fig:spec_sim} a realization of the template with summed residuals
is shown as example (green curve).
\begin{figure}	
	\includegraphics[width=8.5truecm]{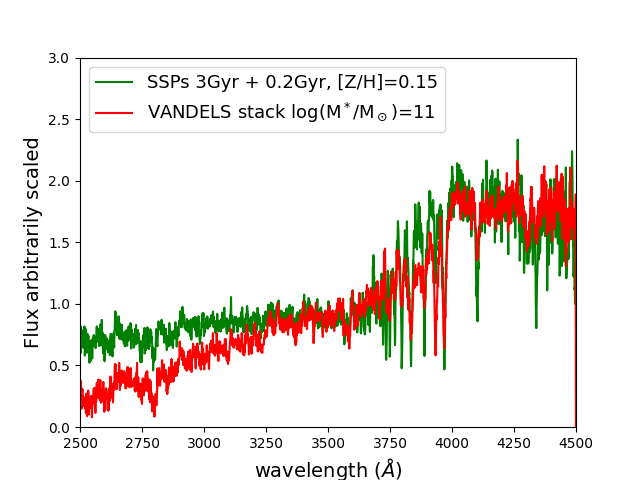}
   \caption{\label{fig:spec_sim} Simulated galaxy with mixed stellar populations.
   The simulated galaxy (green curve) is the weighted sum of a SSP 3 Gyr old weighted 1, and a SSP 0.2 Gyr old weighted 0.1, both with metallicity
   [Z/H]=0.15.
   The VANDELS stacked  spectrum with log(M$^*$/M$_\odot$)=11
   is shown for comparison (red curve).
   The two spectra are normalized to the mean flux in the range [4100-4250] \AA.
}
\end{figure}

Finally, we run \texttt{STARLIGHT} on these simulated spectra with different noise,
for the two spectral ranges [3350-4350] \AA\ and [2600-4350] \AA.
The range [3350-4350] \AA\ provided a mean age of 3.25$\pm$0.2 Gyr
and a mean metallicity [Z/H]=0.11$\pm$0.03;
the range [2600-4350] \AA\ provided a mean age of 3.30$\pm$0.2 Gyr
and a mean metallicity [Z/H]=0.08$\pm$0.02.
Therefore, the presence of a young stellar component, even if accounting
just for about ten per cent of the mass, can induce metallicity estimates
systematically lower (at 1 $\sigma$) when the fit extends to UV range
with respect to the metallicity values derived from optical spectral range.

We remind that this exercise is not meant to probe the nature
of the systematic in the metallicity estimates shown in Sec. 3.2,
since it is based on the assumption that spectral features 
in the UV are well characterized and modeled.
This exercise simply shows that a mixed stellar population could, in principle,
affect in a systematic way the estimate of the mean stellar metallicity when 
the UV spectral range, dominated by younger stellar components and affected
by dust extinction, is taken into account.

We used these simulations also to probe the internal age-metallicity degeneracy
of \texttt{STARLIGHT} full spectral fitting.
This to asses whether and how much the stellar population properties we derived
for VANDELS galaxies can be affected. 
To this end, we searched for a correlation between the age and the metallicity 
values derived for the 30 renditions of the simulated galaxy. 
We did not detect a correlations among the scattered values, meaning that the 
possible systematic is much smaller than the uncertainties due to 
the S/N of our spectra.
}

\section{Extending the spectral fitting to $>$5000 \AA}
In this appendix we check for the possible different stellar population properties
resulting from full spectral fitting performed over different rest-frame wavelength 
ranges.
In particular, a fit performed over [3350-4350] \AA, as in our analysis, and 
a fit extending to $\sim$6000 \AA, as in \cite{carnall22} to verify whether
this is the reason of the discrepancy between our and their stellar metallicity estimates.
We used the high S/N BOSS stacked spectra of massive ETGs at $z$$\sim$0.38 \citep{salvador20}
whose spectral coverage allow us to perform this comparison.
The resulting stellar metallicity values resulting from the \texttt{STARLIGHT} full spectral fitting are shown in Fig. \ref{fig:comp_5850}.
{ The extention of the fitting to $\sim$6000 \AA\ affects the metallicity  
by $\Delta$[Z/H]=-0.04 with respect to the fit performed in the range [3350-4350] \AA. 
This metallicity offset is less than half of the error associated to the metallicity 
estimate of VANDELS stacked spectra and is one seventh of the difference 
($\Delta$[Z/H]$\sim$0.3) beween our 
stellar metallicity estimate for galaxies with log(M$_*$/M$_\odot$)$>$11 and the estimate by \cite{carnall22}.
Therefore, the extension of the fitting to wavelength $\lambda$$>$4350 \AA\ does
not justify the discrepancy between our and their estimate.}

\begin{figure}	
	\includegraphics[width=8.5truecm]{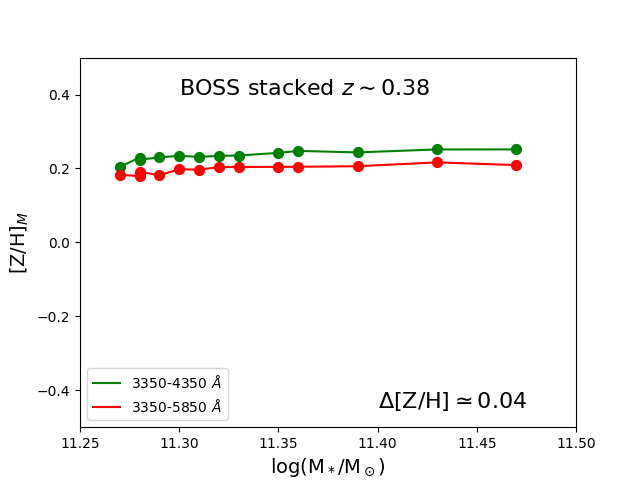}
   \caption{\label{fig:comp_5850} Mass-weighted metallicity of massive ETGs at $z$$\sim$0.38 
   as a function of mass as resulting from the fitting of BOSS stacked spectra 
   \citep{salvador20} over two different wavelength ranges, 3350-4350 \AA\ (green curve) 
   and 3350-5850 \AA\ (red curve). The fitting was performed 
   with \texttt{STARLIGHT} and EMILES models.
}
\end{figure}

\section{Absorption line spectral indices }
\label{sec:indices}
\begin{figure}	
	\includegraphics[width=8.5truecm]{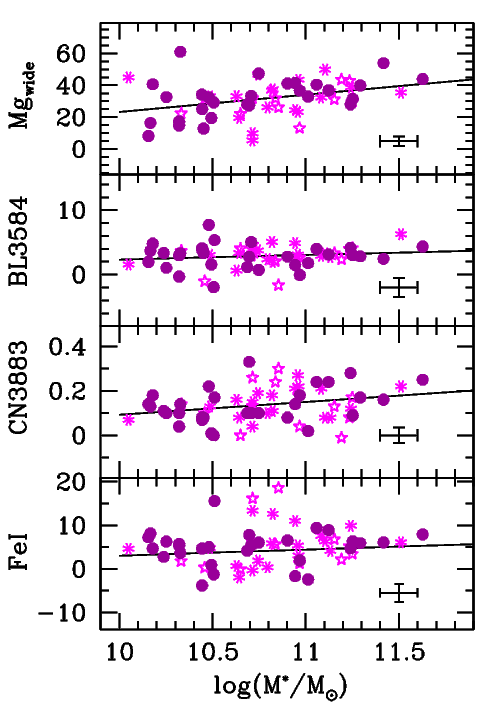}
   \caption{\label{fig:index_mass} The metallicity sensitive indices Mg$_{wide}$,
   Bl3584, CN3883 and FeI (see Table \ref{tab:index}) are shown as a function of 
   the stellar mass of ETGs (purple filled dots), unclassified galaxies (light purple skeletal
   symbols) and S galaxies (stars). The black line in each panel is the
   best-fitting line to the data (orthogonal fit).
   The black crosses on the bottom-right corners represent the typical formal error (based
   on photon statistic) on the data.
  }
\end{figure}


Absorption line indices in the rest-frame wavelength range 2600-4200(4350)\AA\ were measured
for the whole sample of passive galaxies (for those at $z$$<$1.3) and for the stacked galaxies making use of the software \texttt{LECTOR.}\footnote{http://www.iac.es/galeria/vazdekis/vazdekis software.html}
The list of the measured indices and their definition are reported in Tab. \ref{tab:index}.
The strength of the 4000 \AA\ break was measured according to the D4000 definition by
\cite{bruzual83} \citep[see also][]{gorgas99} and the D$_n$ definition by \cite{balogh99}.
The measured indices were corrected for the broadening of the spectra due to the 
low instrumental resolution (R$\sim$600).  
The corrections were obtained by comparing the indices measured on a SSP 3 Gyr old 
(the median age of the sample) smoothed to the instrumental resolution (R$\simeq$600), 
and those of the same model at the nominal resolution of the EMILES spectral library.
We verified that considering ages older or younger than 3 Gyr in the range 1-5 Gyr 
would produce completely negligible variations in the corrections. 
The corrected measured indices with their errors are summarized in Tables \ref{tab:strength}
and \ref{tab:errors}, respectively.

Figure \ref{fig:index_mass} shows the metallicity sensitive features Mg$_{W}$, BL3584,
CN3883 and FeI of ETGs, LTGs and unclassified galaxies as a function of their
stellar mass. 
The black line represent the best-fitting linear relation.
An increasing trend with the mass for the indices considered is visible
{  even if the correlation is not statistically significant given the large scatter}.

\begin{table*}
\begin{minipage}[t]{1\textwidth}
\caption{\label{tab:strength} Absorption line spectral indices (full table available in electronic form).}
\centerline{
\begin{tabular}{lcccccccccccccc}
\hline
\hline
ID	    & MgII  &  MgI &  MgW  & FeI   & CN3883  & CaII(H\&K) & H$\delta_A$ & H$\delta_F$ & CN1  &  CN2   &  Ca4227 & G-b  & D4000 & D$_n$ \\
        & \AA   & \AA  & \AA   &\AA    & mag     &  \AA   & \AA         & \AA         & mag  & mag    & \AA     & \AA  &       &      \\
\hline
UDS021385   &  26.2  &  5.5 & 100.0 &  15.6 &  0.166  & 22.1   & 1.9  &  1.8 &  0.034 &   0.066&   1.4  & 3.6  & 1.48 &  1.24  \\
UDS197769   &  15.8  &  8.4 &  26.1 &	0.4 &	0.096 &  12.2  &  5.4 &  4.4 & -0.081 &  -0.043&   2.0  & 3.0  & 1.38 &  1.15  \\
CDFS126089  &   5.5  &  2.9 &  29.3 &	3.4 &	0.167 &  18.5  &  3.6 &  1.9 & -0.107 &  -0.039&   1.0  &  3.1 & 1.39 &  1.10  \\
\hline
\end{tabular}
}
{The indices are corrected for the broadening of the spectra due to the 
low instrumental resolution (see \S\ 2). 
Values equal to 99.99 indicate failed or unreliable measurements because of problems in the spectrum 
(e.g., uncovered wavelength range, strong emission/absorption sky line).}
\end{minipage}
\end{table*}

\begin{table*}
\caption{\label{tab:errors} Errors (based on photon statistics) on measurements of absorption line indices
(full table available in electronic form).}
\centerline{
\begin{tabular}{lrrrrrrrrrrrrrr}
\hline
\hline
ID	    & eMgII  & eMgI & eMgW & eFeI & eCN3883 & eCaIIHK & eH$\delta_A$ & eH$\delta_F$ & eCN1 & eCN2 &  eCa4227 & eG-b  & eD4000 & eD$_n$ \\
\hline
UDS021385   &  1.6  &  0.8 & 100.0 & 1.6 &   0.03 &   1.3  &  0.3 &  0.2 &  0.02  &  0.03  &   0.1  & 0.4  &  0.04   & 0.03\\
UDS197769   &  0.6  &  0.4 & 1.3 &   1.0 &   0.03 &   1.2  &  0.2 &  0.1 &  0.02  &  0.03  &	0.1  & 0.3 & 0.03	 & 0.03\\
CDFS126089  &  1.2  &  0.7 & 2.1 &   1.5 &   0.03 &   1.4  &  0.3 &  0.2 &  0.02  &  0.03  &	0.2  & 0.4 & 0.03	 & 0.03\\
\hline
\end{tabular}
}
\end{table*}

\begin{table}
\caption{\label{tab:index} Definition of absorption line spectral indices.}
\centerline{
\begin{tabular}{lrrrrc}
\hline
\hline
Index   &   Blue cont.& Feature&   Red cont.&  A/M$^a$ & ref\\
\hline
 Fe2609      & 2562-2588 & 2596-2622 & 2647-2673 &A & 1,2,3,4 \\
 BL2740      & 2647-2673 & 2736-2762 & 2762-2782 &A & 1,2,3,4 \\
 MgII	     &  2762-2782 & 2784-2814 & 2818-2838 & A & 1,2,3,4 \\
 MgI	     &  2818-2838 & 2839-2865 & 2906-2936 & A & 1,2,3,4 \\
 Mg$_{Wide}$ &  2470-2670 & 2670-2870 & 2930-3130 & A & 1,2,3,4 \\
 FeI	     &  2906-2936 & 2965-3025 & 3031-3051 & A & 1,2,3,4 \\
 D4000       &  3750-3950 &           & 4050-4250 &   & 9,10 \\
 D$_n$       &  3850-3950 &           & 4000-4100 &   & 11 \\
 Mg3334      & 3310-3320  & 3328-3340 & 3342-3355 & A & 4, 6 \\
 BL3584      & 3540-3569  & 3570-3600 & 3601-3630 & A &  ...\\
 CN3883       &  3760-3780 & 3780-3900 & 3900-3915 & M & 5 \\
 CaII(H\&K)  &  3900-3915 & 3915-4000 & 4000-4020 & A & 6 \\
 H$\delta_A$ &  4041-4079 & 4083-4122 & 4128-4161 & A & 7,8 \\
 H$\delta_F$ &  4057-4088 & 4091-4112 & 4114-4137 & A & 7,8 \\
 CN1	     &  4080-4117 & 4142-4177 & 4244-4284 & M & 7,8 \\
 CN2	     &  4083-4096 & 4142-4177 & 4244-4284 & M & 7,8 \\
 Ca4227      &  4211-4219 & 4222-4234 & 4241-4251 & A & 7,8 \\
 G-band	     &  4266-4282 & 4281-4316 & 4318-4335 & A & 7,8 \\
\hline
\end{tabular}
}
{$a$ A=EW(\AA), M=magnitude; $1$ - \cite{fanelli90}; $2$ - \cite{chavez07}, $3$ - \cite{maraston09}, 
$4$ - \cite{vazdekis16}, $5$ - \cite{davidge94}, $6$ - \cite{serven05},
$7$ - \cite{worthey97}, $8$ - \cite{trager98}, 9 - \cite{bruzual83,gorgas99}, 11 - \cite{balogh99}.}
\end{table}

\subsection{Age and metallicity $vs$ stellar mass and SFH.}
\begin{figure*}	
	\includegraphics[width=14truecm]{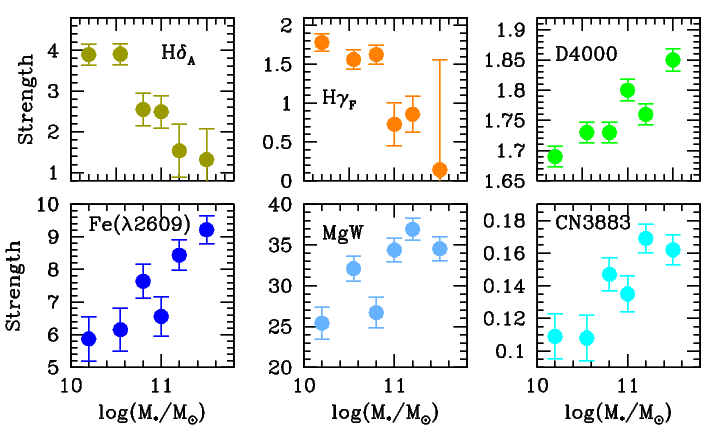}
	\includegraphics[width=12truecm]{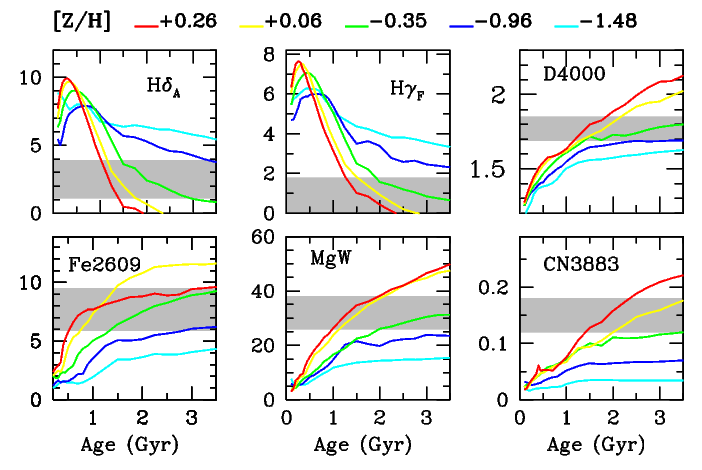}
   \caption{\label{fig:stack_indices} Upper panel - The age sensitive indices H$\delta_A$, 
   H$\gamma_F$ and
   D4000, and metallicity sensitive indices Fe($\lambda$2604), MgW
   and CN3883 (see Table \ref{tab:index}) measured on stacked spectra (large filled symbols)
    are shown as a function of the stellar mass.
   Lower panel - The different curves are the same indices above as predicted by EMILES 
   models for different metallicity values shown as a function of age of the stellar population. The gray regions mark the range of the values of the indices measured on
   stacked spectra.  
   }
\end{figure*}
In this Section, we use the indices to qualitatively verify, in a model independent way, the 
relationships between SFH, age, metallicity and mass of galaxies resulting from the
analysis in \S 4, 5 and 6.
In the upper panel of Fig. \ref{fig:stack_indices}, the age sensitive indices 
H$\delta_A$, H$\gamma_F$ and D4000, and metallicity sensitive indices 
Fe($\lambda$2604), MgW and CN3883 measured on stacked spectra are shown as a 
function of the stellar mass.
In the lower panel, the same indices predicted by EMILES models are shown as a function 
of the age of the stellar population for different metallicity values.
{ It is important to note that all the indices are sensitive to both age and metallicity,
and in particular UV spectral indices are strongly sensitive to even tiny fractions of young stars in a stellar population \citep[see ][]{salvador20}
Therefore, individually, they cannot be used to directly derive the age and the metallicity 
of the underling stellar population.
However, taken together, they provide information on the SFH and the
metallicity independently of models.

The Balmer lines H$\delta_A$ and H$\gamma_F$ are prominent in hot, 
high-mass (1.5-2 M$_\odot$) rapidly evolving stars.
Their main sequence lifetimes are less than 1 Gyr  \cite[e.g.][]{poggianti97}.
Therefore, the strength of these indices are related to the time since the last episode 
of star formation, the weaker the feature the longer the time elapsed since last burst.
This is shown in the lower panel of Fig. \ref{fig:stack_indices} where the 
expected H$\delta_A$ and H$\gamma_F$ indices for EMILES SSPs are shown.
These two indices measured on stacked spectra (upper panel) are anti-correlated
with the mass, suggesting that higher mass galaxies experienced their last burst at 
earlier epochs than lower mass ones, in agreement with the SFH resulting from the 
FSF (see Figures \ref{fig:stack}, and \ref{fig:sfh_cumul}).

The D4000 index \citep{bruzual83} is a measure of the discontinuity produced by 
the opacity of the stellar atmospheres that depends on the ionized metals hence, 
on the stellar temperature.
Hot (high-mass) stars do not contribute to the amplitude of D4000 being their opacity 
low because their elements are ionized.
Therefore, the D4000 index is dominated by low mass stars, its amplitude 
is related to the age { (and metallicity)} of the bulk of the stellar population: 
it gets larger for increasing 
age (and metallicity) as shown in the lower panel of Fig. \ref{fig:stack_indices}.
The D4000 index measured on stacked spectra shows a clear positive trend with the mass 
indicating that the age of the stellar population increases with the  mass of the galaxy, 
in agreement with the results derived from the FSF analysis summarized 
in Fig. \ref{fig:tf_mass} and in Eq. \ref{eq:tf}.
Therefore, the behavior of the Balmer lines, combined with the D4000 index, suggest that 
galaxies with larger stellar masses formed their stars in a shorter SF episode that occurred earlier on in cosmic time.  
This also implies that they have higher SFR. 
This is in agreement with what shown in the lower panel of Fig. \ref{fig:4plots_zh}.

The strength of the indices Fe($\lambda$2604), MgW and CN3883 is sensitive to the 
abundance of those elements in the stellar atmospheres. 
The strength of the three indices (upper panel of Fig.
\ref{fig:stack_indices}) increases systematically with the mass, suggesting a corresponding 
increase of the metallicity\footnote{However, notice that CN3883 is mostly sensitive to C and N abundance ratios, with both elements giving only a minor contribution to the total metallicity of a stellar metallicity.}.
It is interesting to note that the indices predicted by models (lower panel) 
for metallicity lower than solar (green, blue and cyan curves) do not increase 
significantly with age in the range 1-3.5 Gyr, and do not match the observed values, 
especially MgW and CN3883.
The lower indices values measured on the stacked 
spectra (whose range is represented by the 
gray regions in the figure) associated with lower mass galaxies, { seem to rule out metallicity 
values much lower than} [Z/H]$\sim$-0.35.
At the same time, the upper limits associated with high-mass galaxies require metallicity 
higher than solar to be matched.
Analogous conclusions can be reached looking at the age sensitive indices
H$\delta_A$, H$\gamma_F$ and D4000, whose range of values provide a similar
lower limit to metallicity when compared to the indices predicted by models.

In conclusion, the metallicity-sensitive indices that we measured all suggest a positive 
trend of the stellar metallicity with 
the stellar mass of the galaxy, in agreement with the results obtained in Sec. \ref{sec:metal}.
Hence, the metallicity follows the same trend
with the SFH of galaxies shown in Fig. \ref{fig:4plots_zh}.
These results seem to confirm the existence of the stellar mass-metallicity relation of 
passive galaxies at $z\leq1.4$ (cosmic time $<$4.5 Gyr) independently of models and 
of spectral fitting.
This, in turn, implies that the stellar metallicity and the resulting relation have 
been established well in advance, during the formation process.
It is worth noting that, taking as reference EMILES models {at solar abundance ratio}, 
the indices associated to 
higher mass galaxies are consistent with supersolar metallicity, while those associated 
to lower mass galaxies are consistent with lower metallicities.

\subsection{Constraints on [$\alpha$/Fe] ratio and its evolution}
The Mgb[$\lambda$5177] index and Fe lines at $\lambda$$>$5000 \AA\ (Fe5015, Fe5270, and Fe5335)
are considered { the best lines to constrain the [Mg/Fe] abundance ratio} \citep[e.g.][]{trager98, thomas03}. 
However, these features fall outside the rest-frame wavelength range 
covered by VIMOS spectra of VANDELS galaxies at 1.0$<$$z$$<$1.4.
Therefore, we considered the UV spectral features MgW, MgI, MgII, Fe2609 and FeI 
(see Tab. \ref{tab:index} for the indices definition).

UV spectral indices \citep[see e.g.][for a description]{maraston09,vazdekis16} are more 
affected by the presence of young populations than optical indices.
Moreover, the effect of elemental abundance ratios on these indices is not yet known 
and they are still affected by higher uncertainties in the stellar population synthesis models.
For instance, they can be affected by the presence of the UV-upturn, particularly MgII, as discussed by, e.g., \cite{lecras16} and \cite{lonoce20}.

Fig. \ref{fig:alphaFe} shows Magnesium indices versus Iron index for VANDELS galaxies
at $<$$z$$>$$\sim$1.2 and for massive log(M$_*$/M$_\odot$)$>$11 BOSS ETGs at $<$$z$$>$$\sim$0.38 \citep{salvador20}.
As Magnesium tracers we considered MgW and MgI and, for Iron, we defined the quantity 
$<$Fe$>$=(Fe2609+FeI)/2.
{ It is worth to remind that the expected effect of enhancing [$\alpha$/Fe] 
on Mg-Fe diagrams is to make Mg higher and Fe lower, i.e. points should be offset along an oblique direction compared to models predictions for varying age and metallicity}.
In the left panels, individual VANDELS galaxies are colored
according to their stellar mass.
The positive trend with the mass of both Magnesium and Iron confirms,
once again independently of models, the mass-metallicity trend found in \S 4.
In the right panels galaxies are colored according to their $<$SFR$>$ (see \S 6).
Metallicity depends also on $<$SFR$>$, even if the plot MgI-$<$Fe$>$ shows a 
more complex dependence, suggesting that
$<$SFR$>$ may affect mainly Fe.
In fact, Fig. \ref{fig:4plots_zh} shows that also the duration of the SF plays 
an increasingly important role in the metallicity of a galaxy as the mass decreases
(see Discussion).

We probed the possible evolution of [$\alpha$/Fe] ratio in massive 
early-type galaxies by comparing the same indices measured for VANDELS galaxies 
with those measured on the stacked  spectra of massive ETGs at $<$$z$$>$$\sim$0.38.
In the MgW-$<$Fe$>$ plane. 
These massive ETGs are offset with respect to VANDELS stacks
and are located among individual VANDELS galaxies with highest MgW, suggesting a possible
higher [$\alpha$/Fe] abundance. 
{ However, this is not the only possible explanation for the offset.
Indeed, \cite{salvador20} show that it is Fe2609 in their spectra to be too low compared
to SSP model predictions because of the presence of young stellar populations. 
Therefore, it is not necessarily MgW to be too high, but indeed the offset could be
explained because of Fe being too low.}
In the MgI-$<$Fe$>$ plane, low redshift massive ETGs are not offset, they are exactly 
superimposed to VANDELS stacks.
{ Therefore, the comparison between UV indices of VANDELS galaxies at $<$$z$$>$$\sim$1.2 
and massive ETGs at $<$$z$$>$$\sim$0.38 shown in Fig. \ref{fig:alphaFe} does not provide 
a clear answer about a possible redshift evolution of [$\alpha$/Fe] in massive galaxies.}

To constrain the [$\alpha$/Fe] value, in Fig. \ref{fig:alphaFe} the EMILES models
for 3 fixed ages (as in the legend) and for metallicity in the range 
-0.96$\le$[Z/H]$\le$0.26 are shown.
These models assume [Z/H]=[Fe/H] and are based on
the empirical stellar spectra following the Milky Way (MW) abundance
pattern as a function of metallicity.
Therefore, in these models, [$\alpha$/Fe]$\sim0.0$ for metallicity solar or higher, while
at lower metallicity this is not true \citep[see][for a detailed descritpion]{vazdekis16}.
VANDELS galaxies perfectly agree with EMILES models independently of the Magnesium tracer
considered, while massive ETGs at $<$$z$$>$$\sim$0.38 are offset when the MgW index 
is considered.
{ On the basis of these diagnostic plots, and considering the poor knowledge of the effects
of element abundance ratios on UV indices, we cannot draw a firm
conclusion about the evolution of the [$\alpha$/Fe] of massive galaxies from UV indices.}

\begin{figure*}	
	\includegraphics[width=8.5truecm]{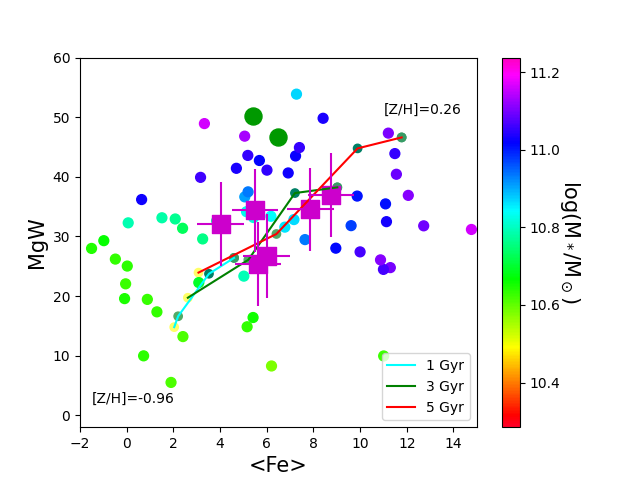}
	\includegraphics[width=8.5truecm]{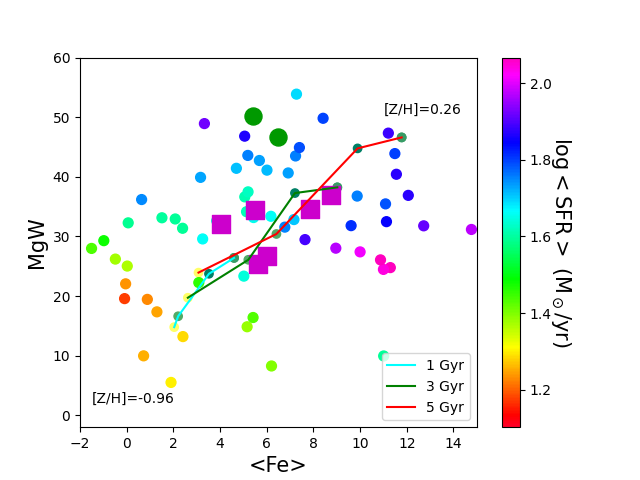}
 	\includegraphics[width=8.5truecm]{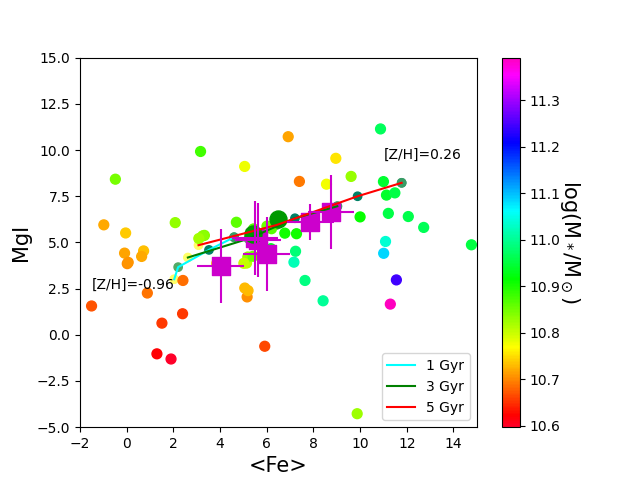}
	\includegraphics[width=8.5truecm]{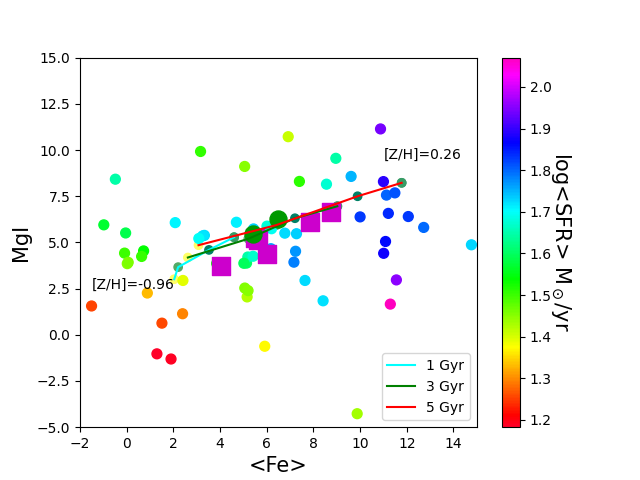}
   \caption{\label{fig:alphaFe} Magnesium versus Iron line indices for 
   individual VANDELS galaxies (small colored filled circles), VANDELS stacks (magenta
   big squares) and BOSS stacks of massive ETGs \citep[big green filled circles;][]{salvador20}.
   These latter are shown as two points representing the mean values obtained from the
individual stacks in the two velocity dispersion intervals [160-220] km/s and [220-280] km/s
respectively.
   In the upper panels the Magnesium index MgW  is plotted versus 
   the index $<$Fe$>$ defined as $<$Fe$>$=(Fe2609+FeI)/2.
   In the lower panels the Magnesium index MgI
   is plotted versus $<$Fe$>$ (see Tab. \ref{tab:index} for the indices definition). 
   In the left panels, individual VANDELS galaxies are colored according to their stellar mass,
   in the right panels according to their $<$SFR$>$ (see \S 6).
   Errorbars on VANDELS stacks are the MAD of the values of individual galaxies.
   Lines show the predictions of EMILES models for 3 fixed ages (as in the legend) 
   for metallicity in the range -0.96$\le$[Z/H]$\le$0.26 and [$\alpha$/Fe]$\sim$0.0
   (see text).
   }
\end{figure*}



\bsp	
\label{lastpage}
\end{document}